\documentstyle[fleqn]{elsart}
\begin{document}
\renewcommand{\thesection}{\arabic{section}}
\renewcommand{\theequation}{\thesection.\arabic{equation}}
\newcommand{\ezero}{\setcounter{equation}{0}}
\newcommand{\ds}{\displaystyle}
\newcommand{\nl}{\nonumber \\}
\newcommand{\ba}{\vspace{-.2cm} \begin{eqnarray}}
\newcommand{\ea}{\end{eqnarray}}
\newcommand{\beq}{\vspace{.2cm}\begin{equation}}
\newcommand{\eeq}{\end{equation}}
\newcommand{\xsec}{cross-sec\-tion}
\newcommand{\xsecs}{cross-sec\-tions}
\newcommand{\dis}{distri\-bu\-tion}
\newcommand{\diss}{distri\-bu\-tions}
\newcommand{\SV}{soft+virtual}
\newcommand{\oal}{${\cal O}(\alpha)$}
\newcommand{\zz}{$Z$}
\newcommand{\wpl}{$W^+$}
\newcommand{\wmi}{$W^-$}
\newcommand{\epl}{$e^+$}
\newcommand{\emi}{$e^-$}
\newcommand{\ee}{\epl\emi}
\newcommand{\sm}{{\tt SM}}
\newcommand{\itm}{\item[$\bullet$]}
\newcommand{\sone}{$s_1$}
\newcommand{\stwo}{$s_2$}
\newcommand{\ZZ}{Z^0}
\newcommand{\WPL}{W^+}
\newcommand{\WMI}{W^-}
\newcommand{\EPL}{e^+}
\newcommand{\EMI}{e^-}
\newcommand{\EE}{\EPL\EMI}
\newcommand{\SONE}{s_{1}}
\newcommand{\STWO}{s_{2}}
\newcommand{\SLAM}{\sqrt{\lambda}}
\newcommand{\SPR}{s'}
\newcommand {\oalf} {\mbox{${\cal O}(\alpha)$}}
\newcommand{\nn}{\noindent}
\newcommand{\nll}{\nonumber \\}
\newcommand{\hf}{\hfill}
\newcommand{\bq}{\begin{equation}}
\newcommand{\eq}{\end{equation}}
\newcommand {\oa}{\mbox{${\cal O}(\alpha)$}}
\newcommand{\mr}{\mathrm}
\newcommand{\ltwo}{LEP~2}
\newcommand{\lone}{LEP~1}
\newcommand{\SM}{\cal{SM}}
\newcommand{\ww}{\mbox{$\Gamma_W$}}
\newcommand{\nobody}{\rule{0ex}{1ex}}
\newcommand{\hnobody}{\rule{1ex}{0ex}}
\newcommand{\vc}{\char'24\hspace{-1ex}c}
%
%
%
\begin{frontmatter}

\begin{flushleft}
DESY 96--233
\\
hep-ph/9612409
\\
December 1996
\end{flushleft}

\title{{\tt GENTLE/4fan v. 2.0} 
\\
A Program for the Semi-Analytic 
Calculation of Predictions for the Process $e^+e^- \to 4f$
}
\author{D. Bardin$^{a,1,2}$, J. Biebel$^a$, D. Lehner$^{b,3}$,}
\author{A. Leike$^{c,4}$, A. Olchevski$^d$, T. Riemann$^{a,2}$}
\address[a]{Deutsches Elektronen-Synchrotron DESY,
         Institut f\"ur Hochenergiephysik IfH~Zeuthen,
         Platanenallee 6, D-15738 Zeuthen, Germany}
\address[b]{Fakult\"at f\"ur Physik, Albert-Ludwigs-Universit\"at,
         Hermann-Herder-Str.~3, D-79104 Freiburg, Germany}
\address[c]{Sektion Physik, Ludwig-Maximilians-Universit\"at,
            Theresienstra{\ss}e 37, D-80333~M\"unchen, Germany}
\address[d]{Laboratory of Nuclear Problems, JINR, ul. Joliot-Curie 6,
            RU-141980 Dubna, Moscow Region, Russia}
\footnotetext[1]{On leave of absence from Bogoliubov Theor. Lab., JINR,
                 ul. Joliot-Curie 6, RU-141980 Dubna, Moscow Region,
                 Russia}
\footnotetext[2]{Supported by European Union with grant INTAS--93--744}
\footnotetext[3]{Address after 1 August 1996: INNOSOFT Ltd.,
                 Bodenseestra{\ss}e 4, D-81241 M\"unchen, Germany}
\footnotetext[4]{Supported by European Union with grant CHRX--CT940579}

\vfill

\begin{abstract}
  We describe version 2.0 of the {\sc Fortran} program {\tt GENTLE/4fan}
  for the semi-analytic computation of \xsecs\ and \diss\ in
  four-fermion production in \ee\ annihilation.
  {\tt GENTLE/4fan} covers all charged current and neutral current
  four-fermion final states with no identical particles, no electrons,
  and no electron neutrinos in the final state.
  Initial state radiation representing the most relevant quantum
  corrections and anomalous triple gauge boson couplings have been
  included.
\end{abstract}
\vfill
\end{frontmatter}
%
%
\newpage

\section*{PROGRAM SUMMARY}
   {\em Title of program\/}:  {\tt GENTLE/4fan}
\\ {\em Version\/}:           2.0 (December 1996)
\\ {\em Catalogue number\/}:
\\ {\em Program obtainable from\/}: 
\\
http://www.ifh.de/theory or on
request by e-mail from the authors
\\ {\em Licensing provisions\/}: non
\\ {\em Computers\/}: all
\\ {\em Operating system\/}:  all
\\ {\em Program language\/}:     {\tt FORTRAN-77}
\\ {\em Memory required to execute with typical data\/}: 0.5 Mb
\\ {\em No. of bits per word\/}: 32 
\\ {\em No. of lines in distributed program\/}:  8300
\\ {\em Other programs called\/}: none
\\ {\em External files needed\/}: none
\\ {\em Keywords\/}:  $e^+e^-$ annihilation, QED, higher order corrections,
\\  four-fermion production, $WW,ZZ,Z\gamma,\gamma\gamma,ZH$ production
\\  {\em Nature of physical problem\/}:
\\  Description of two-boson production in  $e^+e^-$ annihilation at
high energies with selected possible background
contributions; initial state QED corrections (ISR); selected final state
corrections; anomalous triple gauge boson couplings
\\   {\em Method of solution\/}:
\\ Semi-analytic; numerical integration of analytic formulae:
\\ (i) Born approximation: 
Numerical integrations over two virtualities $s_1,s_2$
plus optionally over the boson production angle $\theta$;
\\ (ii) QED corrections in the flux function approach (FF):
one additional numerical integration over virtuality $s'$;
\\ (iii) QED corrections in the structure function approach (SF):
alternatively to (ii), two additional numerical integrations over $x_1,x_2$ 
\\ {\em Restrictions on complexity of the problem\/}:
\\ Only selective experimental cuts are possible;
   the program calculates total cross-sections and the distribution in
the boson production angle; 
background contributions are available only for selected final state
configurations; no inclusion of pure weak corrections 
\\ {\em Typical running time\/}:
\\ The running time strongly depends upon the options used;
   one finds e.g.: 
\\ Born cross-section on-shell: below one sec., off-shell: few secs.;
\\ LLA QED corrections (FF) on-shell: few secs.; off-shell: few mins.;
\\ LLA QED corrections (SF) on-shell: few secs.; off-shell: several mins.;
\\ complete ISR (FF): on-shell: few mins.,
off-shell: several hours;
\\ with background: from few secs. (no QED) to several hours (with QED);
\\ with anomalous couplings: from few secs. (no QED) to several
mins. (with QED) 
\newpage
\tableofcontents
\newpage
\section*{LONG WRITE-UP}
\vspace{.5cm}
\section{Introduction}
\label{intro}
\ezero
With the advent of LEP2, now running above the $W$ pair threshold, \ee\
collisions have entered their multiple heavy boson production era which
will continue with a higher energy linear \ee\ collider (LC).
Among the most interesting processes to be studied in high energy \ee\
annihilation is boson pair production.
As heavy bosons decay, their physics is only accessible through their
decay products.
Thus, at LEP2 or LC, boson pair production must be accessed through
the experimental analysis of four-fermion final states,
\beq
  \EE \rightarrow (W^+W^-, ZZ, Z\gamma, \gamma \gamma, ZH, \ldots)
  \rightarrow f_1 {\bar f}_2 f_3 {\bar f}_4~~.
  \label{eeto4f}
\eeq
%
%
\section{Physics Background}
\label{physback}
\ezero
The Standard Model (\sm) four-fermion final states produced in \ee
annihila\-tion have been classified by several
authors~\cite{LEP2MC,LEP2SM,teupitz94,berendsclass}.
Using the semi-analytic approach, we have so far studied the production
of four-fermion final states that contain neither identical particles
nor electrons or electron neutrinos.
These four-fermion final states are printed in {\bf boldface} in
tables~\ref{cctype} and~\ref{nctype}.
\begin{table}[ht]
\begin{center}
\vspace{.3cm}
\caption{\it Number of Feynman diagrams for `{\tt CC}' type final states.
         \label{cctype} } \vspace{.25cm}
\begin{tabular}{|c|c|c|c|c|c|} \hline
  & \raisebox{0.pt}[2.5ex][0.0ex]{${\bar d} u$}
  & ${\bar s} c$ & ${\bar e} \nu_{e}$
  & ${\bar \mu} \nu_{\mu}$ & ${\bar \tau} \nu_{\tau}$ 
  \\ \hline
  $d {\bar u}$ & {\it 43} & {\bf 11} & 20 & {\bf 10} & {\bf 10}
  \\ \hline
  $e {\bar \nu}_{e}$       &  20 &  20 & {\it 56} & 18 & 18
  \\ \hline
  $\mu {\bar \nu}_{\mu}$ & {\bf 10} & {\bf 10} & 18 & {\it 19} & {\bf 9}
  \\ \hline
\end{tabular}
\vspace{.1cm}
\end{center}
\end{table}
\begin{table}[ht]
\begin{center}
\caption{\it Number of Feynman diagrams for {\tt NC} type final states.
         \label{nctype} } \vspace{.25cm}
\begin{tabular}{|c|c|c|c|c|c|c|c|c|c|} \hline
  & \raisebox{0.pt}[2.5ex][0.0ex]{${\bar d} d$}
  & ${\bar s} s, {\bar b} b$ & ${\bar u} u$ & ${\bar e} e$
  & ${\bar \mu} \mu$ & ${\bar \tau} \tau$
  & ${\bar \nu}_{e} \nu_{e}$ & ${\bar \nu}_{\mu} \nu_{\mu}$
  & ${\bar \nu}_{\tau} \nu_{\tau}$
  \\ \hline
  \raisebox{0.pt}[2.5ex][0.0ex]{${\bar d} d$}
  & {\tt 4$\cdot $16} & {\bf 32} & {\it 43} & {48} & {\bf 24} 
  & {\bf 24} & 21 & {\bf 10} & {\bf 10}
  \\ \hline
  \raisebox{0.pt}[2.5ex][0.0ex]{${\bar u} u$}
  & {\it 43} & {\it 43} & {\tt 4$\cdot$16} & {48} & {\bf 24} & {\bf 24}
  & {21} & {\bf 10} & {\bf 10}
  \\ \hline
  \raisebox{0.pt}[2.5ex][0.0ex]{${\bar e} e$}
  & {48} & {48} & {48} & \sf{4$\cdot$36} & {48} & {48} & {\it 56} & {20}
  & {20}
  \\ \hline
  \raisebox{0.pt}[2.5ex][0.0ex]{${\bar \mu} \mu$}
  & {\bf 24} & {\bf 24} & {\bf 24} & {48} & {\tt 4$\cdot$12} & {\bf 24}
  & {19} & {\it 19} & {\bf 10}
  \\ \hline
  \raisebox{0.pt}[2.5ex][0.0ex]{${\bar \nu}_e \nu_{e}$}
  & {21} & {21} & {21} & {\it 56} & {19} & {19} & \sf{4$\cdot$9} & {12}
  & {12}
  \\ \hline
  \raisebox{0.pt}[2.5ex][0.0ex]{${\bar \nu}_{\mu} \nu_{\mu}$}
  & {\bf 10} & {\bf 10} & {\bf 10} & {20} & {\it 19} & {\bf 10} & {12}
  & {\tt 4$\cdot$3} & {\bf 6}
  \\ \hline
\end{tabular}
\end{center} 
\end{table}        
%
%
\subsection{Off-shell boson pair production}
For semi-analytic tree level calculations (flag {\tt IBORNF=0}), the 
four-particle phase space
is para\-metrized by the boson scattering solid angle, the boson
decay solid products' angles in the corresponding boson rest frames,
and the two 
final state fermion pair invariant masses \sone\ and
\stwo~\cite{cc11,nuniall}.
After analytic integration of all angular variables, the total \xsec\
for 
double-resonant four-fermion production is given by
\beq
  \sigma^{\rm res}(s) \; = \;
  \int \d \SONE \rho_B(\SONE)
  \int \d \STWO \rho_B(\STWO)
  \;\; \sigma_{0}(s;\SONE,\STWO)
  \label{sigres}
\eeq
with the Breit-Wigner densities
\ba\nonumber \\
  \rho_B(s_{i}) \; = \; \frac{1}{\pi} \frac {\sqrt{s_i}  \Gamma_B}
   {|s_{i} - M_B^2 + i s_{i} \, \Gamma_B / M_B |^2}
  \times {\mathrm {BR}} \;\;\,
  \stackrel {\Gamma_B \rightarrow 0} {\longrightarrow} \;\;\,
  \delta(s_{i} - M_B^2) \times {\mathrm {BR}}
\nll
  \label{rhow}
\ea
of the resonant bosons. 
Explicit formulae for the twofold differential
\xsecs\ in the double-differential approximation
$\sigma_{0}(s;\SONE,\STWO)$ are found in reference~\cite{muta}
for $W$\ pair production, in~\cite{teupitz94} for $Z$ pair production,
and in~\cite{nc24h} for associated Higgs production.
We emphasize that these twofold differential \xsecs\ are given by very
simple formulae.
The Breit-Wigner functions are smoothed by mapping (see
appendix~\ref{integration}). 
Flag {\tt IMAP} allows for the {\tt CC} case to use the
Breit-Wigner functions directly.
In the latter case the program works slower by about a factor of two.
We note that in this case the invariant mass distributions are
accessible.
With flag {\tt INCPRC} one may select $ZZ$, $\gamma \gamma$, or their
common production. 
For the {\tt CC} case, the on-shell limit may be calculated with flag {\tt
  IONSHL=0}.

The realization of the various generic functions mentioned here and in
the following sections is described in section~\ref{prgdesc}. 

\subsection{Background contributions}
Although double-resonant amplitudes are dominant in most relevant cases,
single- and non-resonant `background' amplitudes contribute
non-negligible to the \xsecs.
Including background Feynman diagrams, one can use the numbers from
tables~\ref{cctype} and~\ref{nctype} to introduce a transparent
notation~\cite{teupitz94}:
\begin{list}{}{}
  \item[CC] Processes with final states of type
            $f_1^u{\bar f}_1^d f_2^u {\bar f}_2^d$ are called
            {\em charged current processes}:
            {\tt CC11}, {\tt CC10}, {\tt CC09}; {\tt CC20}, {\tt CC18}.
  \item[NC] Processes with final states of type
            $f_1{\bar f}_1 f_2 {\bar f}_2$ are called
            {\em neutral current processes}:
            {\tt NC32}, {\tt NC24}, {\tt NC10}, {\tt NC06}; 
            {\tt NC48}, {\tt NC20}, {\tt NC21}, {\tt NC19};
            {\tt NC4$\cdot$16}, {\tt NC4$\cdot$12}, {\tt NC4$\cdot$03},
            {\tt NC12};
            {\tt NC4$\cdot$36}, {\tt NC4$\cdot$09}. 
  \item[mix] Processes which may be considered as both {\tt CC} and NC types
            are called {\em mixed processes}:
            {\tt mix43}, {\tt mix19}; {\tt mix56}. 
            An example for a mixed process is the production of
            $u{\bar d} d{\bar u} \equiv u{\bar u} d{\bar d}$.
\end{list}
In the {\tt NC} and mix classes, also Feynman diagrams with Higgs bosons
contribute.
In addition to the two-boson production classes {\tt CC03}~\cite{muta}
($W$\ pair production), {\tt NC02}~\cite{nuniall} ($Z$\ pair production),
and {\tt NC08}~\cite{nuniall} ($ZZ, Z\gamma, \gamma \gamma$\ production),
{\tt GENTLE/4fan} is able to perform calculations for {\tt CC11},
{\tt CC10}, and {\tt CC09}~\cite{cc11} as well as {\tt NC32},
{\tt NC24}, {\tt NC10}, {\tt NC06}~\cite{nc24}.
This is chosen by flag {\tt IPROC}.
In the above {\tt NC} cases, also Higgs diagrams are included~\cite{nc24h}.

Complete tree level four-fermion production \xsecs\ are given by the
generic formula
\beq
  \sigma^{\rm{Born}}(s) \; = \; \int \d \SONE \int \d \STWO \;\;
   \frac{\sqrt{\lambda}}{\pi s^2} 
   \sum_k \frac{\d^2 \sigma_k(s;\SONE,\STWO)}{\d \SONE \d \STWO}
  \label{totsig}
\eeq
with $\lambda \equiv \lambda(s;\SONE,\STWO),~\lambda(a,b,c) =
a^2\!+\!b^2\!+\!c^2\!-\!2ab\!-\!2ac\!-\!2bc$ and 
\beq
  \frac{\d^2 \sigma_k}{\d \SONE \d \STWO} \; = \;
  {\cal C}_k(s;\SONE,\STWO)  {\cal G}_k(s;\SONE,\STWO)~~.
  \label{diffsig}
\eeq
The coefficient functions ${\cal C}_k$ represent coupling constants and
boson propagators, while ${\cal G}_k$\ is a kinematical
function obtained after analytical performance of angular integration.
The index $k$ labels cross-section contributions with both different
coupling structure and different Feynman topology.
Noticing that often different twofold differential \xsecs
$\d^2 \sigma_k / \d \SONE \d \STWO$\ are related by symmetries in their
arguments, one realizes that equation~(\ref{totsig}) enables efficient
coding of the \xsec. For {\tt CC} processes,
background contributions are switched off with flag {\tt IBCKGR=0}.
With flag {\tt ICHNNL} one may choose the final state topology in the  
{\tt CC} case and with {\tt IFERM1} and {\tt IFERM2} in the {\tt NC} case.

%
%
\subsection{Initial state QED radiative corrections}
\label{qed}
In \ee\ annihilation, the most relevant radiative corrections are initial
state QED corrections (ISR).
The phase space is now parameterized by the center-of-mass photon
scattering solid angle, by the boson scattering solid angle in the
two-boson (or, equivalently, four-fermion) rest frame, by the boson
decay products' solid angles in the corresponding boson rest frames,
by the 
final state fermion pair invariant masses, and by the four-fermion
(or reduced) invariant mass $s'$~\cite{nuniall}.
After analytic integration over all angular degrees of freedom, the
\oal\ ISR-corrected total four-fermion production \xsec\ with soft
photon exponentiation is determined in the flux function approach
(flag {\tt ICONVL=0}) by
\beq
  \sigma^{\rm{ISR}}(s) \; = \;
  \int \d \SONE \int \d \STWO \;\; \int \d s' \;\;
  \sum_k \frac{\d^3\sigma_k(s,s';\SONE,\STWO)}{\d \SONE \d \STWO \d s'}~~,
  \label{ISRxstot}
\eeq
with
\beq
  \frac{\d^3\sigma_k(s,s';\SONE,\STWO)}{d \SONE \d \STWO \d s'} \; = \;
  {\cal C}_k  
  \left[ \beta_e v^{\beta_e - 1} {\cal S}_k+{\cal H}_k \right]~~,
  \label{ISRxsdif}
\eeq
where $\beta_e = 2\,(\alpha/\pi) \, (L - 1)$,
$L \equiv \ln (s/m_e^2)$.
The \SV\ and hard corrections ${\cal S}_k$ and ${\cal H}_k$ separate
into a universal, factorizing, process-independent and a non-universal,
non-factorizing, process-dependent part. 
The latter is included with flag {\tt ITNONU=1}.
They are given by~\cite{nuniall,nunicc}
\ba
\nonumber\\
  {\cal S}_k 
& = &
  \left[1 + {\bar S}(s) \right] \sigma_{k,0}(s';\SONE,\STWO)
  + \:\sigma_{{\hat S},k}(s';\SONE,\STWO) ~~,
  \nonumber \\
  \nonumber \\
  {\cal H}_k
& = &
  \underbrace{{\bar H}(s,s') \:
  \sigma_{k,0}(s';\SONE,\STWO)\;\;\;\;}_{Universal~Part}
  + \underbrace{\sigma_{{\hat H},k}(s,s';\SONE,\STWO)}_{Non-universal~Part}
  \label{softhard}
\ea
with $\sigma_{k,0}(s';\SONE,\STWO) \! \equiv \!
\left[ \sqrt{\lambda}/(\pi s'^{2})\right] \!  
{\cal G}_k(s';\SONE,\STWO)$.
The universal \oal\ \SV\ and hard radiators
${\bar S}$\ and ${\bar H}$ read
\ba
\nonumber \\
  {\bar S}(s) & \: = \: & \frac{\alpha}{\pi}
       \left( \frac{\pi^2}{3} - \frac{1}{2} \right)
     + \frac{3}{4} \, \beta_e~~,
   \label{svrad} 
\\
\nonumber \\ 
  {\bar H}(s,s') & \: = \: & - \frac{1}{2}
       \left(1+\frac{s'}{s}\right)\beta_e~~.
   \label{hardrad}
\ea
Flag {\tt IZERO} influences ${\bar S}(s)$.
If the index $k$ labels pure s-channel $e^+e^-$ annihilation, the
non-universal contribution vanishes.
Non-universal ISR contributions are analytically very complex, but,
in contrast to the universal corrections, not
logarithmically enhanced.

For the {\tt CC03} process, a significant QED radiative correction is
the so-called Coulomb singularity~\cite{coulombdb,coulombdur,coulombdurold}
originating from the exchange of photons between the two resonant $W$
bosons.
It effects in a factor $(1+{\cal F}_{Coulomb})$ to be multiplied to the
twofold differential \xsec\ of equation~(\ref{diffsig}) or to the
threefold differential \xsec\ of equation~(\ref{ISRxsdif}). 
Details are defined with flag {\tt ICOLMB}.

The leading logarithmic ISR contributions (essentially what we called
the universal ISR contributions) may also be determined using the
so-called structure function approach~\cite{LEP2WW} (flag {\tt
ICONVL=1}). 
The basic formula, adapted for our case, is 
\beq
  \frac{\d^2 \sigma_{\EE\to4f\,n\gamma}}{\d \SONE \d \STWO}
  \; = \;
  \int_{x_+^{\min}}^1 \d x_+  \int_{x_-^{\min}}^1 \d x_- \;
  D(x_+) D(x_-) \: \sigma_0(s';\SONE,\STWO)
  \label{strucxs}
\eeq
with the tree level twofold differential \xsec\ $\sigma_0$ evaluated at
$s'=s x_+ x_-$.
The variables $x_+$\ and $x_-$\ represent the momentum fractions of
the \epl\ and \emi\ after the radiation of photons.
Their minimum values are given by
\ba
\nonumber \\
  x_+^{\min} &=& \frac{(\sqrt{s_1}+\sqrt{s_2})^2}{s}, 
\nl
\nonumber \\
  x_-^{\min} &=& \frac{(\sqrt{s_1}+\sqrt{s_2})^2}{x_+ s}.
\ea
An exhaustive discussion of the application of the structure function
approach to $4f$ production may be found in~\cite{LEP2MC,LEP2WW} and
in references quoted therein. 
Our implementation of the structure functions $D(x)$ 
follows equations~(4.18) to~(4.21) of reference~\cite{cc11}.
Note that, to improve the calculation, $\sigma_{0}$\ can be replaced by
a ``dressed'' tree level \xsec\ which includes the running QED
coupling and/or the Coulomb factor.
Since the soft-photon pole is proportional to $(L-1)$ rather than to the
leading logarithm $L$ alone there are the so-called {\tt BETA} and
{\tt ETA} alternatives for the implementation of the structure
functions~\cite{LEP2WW}.
They are selected by flag {\tt IZETTA}.

For both descriptions of QED corrections, the details of higher order
terms are defined by flag {\tt IQEDHS}.

In addition to the total \xsec , {\tt GENTLE/4fan} can compute 
the angular distribution, bin-integrated cross-sections, and moments
for several other physically relevant quantities. These quantities are
selected with the variables {\tt IDCS} and {\tt IREGIM}, the latter
being chosen with flags {\tt IRMAX, IRSTP}: 
\begin{enumerate}
  \item[(1)] the radiative final state four-fermion invariant mass loss
        $X = (s - s') / (2 \sqrt{s}) $\ for the {\tt CC11}
        family in the flux function as well as the structure function
        approach;
  \item[(2)] the true radiative energy loss $X = \sqrt{s}/2
        \left[(1-x_+) + (1-x_-) \right]$\ for the {\tt CC11}
        family in the structure function approach;
  \item[(3)] the $W$ mass shift
        $X = \left[ \left( \sqrt{\SONE} + \sqrt{\STWO} \right)/2 - M_W
      \right]$\ 
        for the {\tt CC11} family in the flux function as well as the
        structure function approach;
  \item[(4)] moments $X$ related to the angular distribution are
    treated 
    in section~\ref{angular}. 
\end{enumerate}
The running of loop variable {\tt IMOMN}=$n$ is defined by flags {\tt
  IMMIN, IMMAX}.
It is used for the selection of the 
moment of $n^{th}$\ degree $M_{(X,n)}$\ of a quantity $X$ (calculated
with QED corrections):
\beq
  M_{(X,n)} = \frac{\int X^n   \sigma_X}{\int \sigma_X}
\label{invmoments}
\eeq
where $X$ is selected by {\tt IREGIM} and  
the integration is over the phase space variables upon which $X$\
depends and $\sigma_X$\ is the \xsec\ differential in exactly those
phase space variables.
Note that $M_{(X,1)}$ is an average of the quantity $X$.

Finally, we mention that an inclusive
implementation of QCD corrections is provided.
Firstly, QCD corrections to the total $W$ width have been implemented
by a 
factor $\left(1 + 2\alpha_s/3\pi \right)$ to be multiplied
to $\Gamma_W$.
Secondly, the final state radiative QCD corrections to the formation
of a quark-anti-quark pair are built in by multiplying
the relevant branching ratio with $\left(1 + \alpha_s/\pi
\right)$. 

%
%
\subsection{Angular distributions}
\label{angular}
Because of the importance of the {\tt CC03} process, {\tt GENTLE/4fan}
is also able to compute $W$ production angular distributions,
bin-integrated cross-sections and moments for this process.
A selection is possible by the flags {\tt IDCS} and {\tt IREGIM}.
In the first and last of the three mentioned cases, the last
analytical integration is not carried out. 
Instead, in Born approximation the analytic threefold differential \xsec\
\beq
  \frac{\d^3\sigma_{\tt CC03}}{\d \SONE \d \STWO \d\cos\theta}  \; = \; 
  \frac{\sqrt{\lambda}}{\pi s^2} 
  \sum_{k=1}^3 {\cal C}^{\tt CC03}_k 
               {\cal G}^{\tt CC03}_k (s,\STWO,\SONE;\cos\theta)
\label{sigstc}
\eeq
is used. The angular kinematic functions
${\cal G}^{\tt CC03}_k (s,\STWO,\SONE;\cos\theta)$ were published
in~\cite{cc11}.
The \xsec\ $\d \sigma_{\tt CC03}/\d\cos\!\theta$\ is obtained upon
integration over \sone\ and \stwo.
In {\tt GENTLE/4fan}, as a measure for the {\tt CC03} angular
distribution, the first four moments of the differential
\xsec\ (\ref{sigstc}) may be calculated as its convolution with
Chebyshev polynomials $T_n (\cos\!\theta)$. 
The Born moment of $n^{th}$ degree is given by
\beq
  M_{\left(\cos\!\theta, n\right)} \; = \;
  \int \d\SONE \, \d\STWO \, \d\cos\!\theta \; T_n (\cos\!\theta) \,
  \frac{\d^3\sigma_{\tt CC03}}{\d \SONE \d \STWO \d\cos\!\theta}~~.
  \label{angmoments}
\eeq
The zeroth moment is just the total \xsec.

Initial state QED corrections to the angular distributions and moments
have to be calculated in the structure function approach.
With photon emission, the
$W$ production angle in the laboratory system,
$\theta_{\rm lab}$, and in the center-of-mass system of the four final
state fermions, $\theta$, are different.
The Lorentz boost relating them depends on both $x_+$ and $x_-$.
The QED corrected angular moments with ISR are given by
\ba
\nonumber\\
  M_{\left(\cos\!\theta_{\rm{lab}}, n\right)} & \, = \, &
  \int \!\!\d\SONE  \int \!\!\d\STWO \int\!\!\d x_+  \int\!\!\d x_-
  \int \!\!\d\cos\!\theta \; 
\nl
  & & 
\times
  D(x_+) D(x_-) \, T_n [\cos\!\theta_{\rm{lab}}(\cos\theta)] \;
  \frac{\d^3\sigma_{\tt CC03}}{\d \SONE \d \STWO \d\cos\!\theta}~.
  \label{angmomqed}
\ea
%
With flags {\tt IDCS, IBIN, IBINNU}, the user may select the
calculation of the angular distribution:
\ba
\nonumber\\
  \frac{\d\sigma_{\tt CC03}}{\d\cos\!\theta_{\rm{lab}}}
  & \; = \; &
  \int \!\!\d\SONE  \int \!\!\d\STWO \int\!\!\d x_+  \int\!\!\d x_-
\nl &&\times~
  \; D(x_+) D(x_-) \:
\sum_i \left|\frac{\partial \cos\!\theta_i}{\partial
  \cos\!\theta_{\rm{lab}}}\right| 
  \frac{\d^3\sigma_{\tt CC03}}{\d \SONE \d \STWO \d\cos\!\theta_i} 
  \label{angdiffqed}
\ea
with the threefold differential tree level \xsec\ 
$\d^3\sigma_{\tt CC03}/(\d \SONE \d \STWO \d\cos\!\theta_i)$\ depending on
$s'=s x_+ x_-$, \sone, \stwo, and
$\cos\!\theta_i(s,s_1,s_2,x_+,x_-,\cos\!\theta_{\rm lab})$.
Index $i$ indicates that there may be zero, one, or two solutions for
the Lorentz boost.

In analogy to~(\ref{angmomqed}),
the bin-integrated cross-section formulae may be written with the
boosted angular argument under the integral, thus allowing for
relatively simple analytic angular integrations (and avoiding the
appearance of the Jacobean) if, of course, the integration limits set
in the laboratory frame are transformed into angular limits in the
boosted frame.
In this respect, the approach to the bin-integrated cross-section
differs from that for the angular distribution where we integrate
over the laboratory angle (and have to take into account the
Jacobean).
A choice can be made with flag {\tt IBIN}.  

%
%
\subsection{Anomalous couplings}
Angular distributions as defined in the above section~\ref{angular}
are especially useful in the context of anomalous triple boson
couplings~\cite{LEP2ano}.
In {\tt GENTLE/4fan}, only those anomalous $\gamma\WPL\WMI$\ and
$Z\WPL\WMI$ 
couplings are implemented which 
obey Lorentz invariance, conserve {\sf CP}, and do not
modify the electromagnetic interaction.
These constraints allow a restricted set of anomalous contributions to
the triple gauge boson vertex.
These couplings are denoted by $g_3^a(V)$\ with
$V\!\in\!\{\gamma,Z\}$.
The $a = s, x, y, z$ denotes the coupling type including the \sm\
triple-boson coupling $g_3^s(V)$. 
The couplings are 
derived from the anomalous couplings
defined in equations~(\ref{arrbeg}) to~(\ref{arrend}).
The threefold differential {\tt CC03} \xsec\ is then given by~\cite{biebel}
\ba
  \frac{\d^3\sigma^{ano}_{\tt CC03}}{\d \SONE \d \STWO
    \d\cos\!\theta_{\rm{lab}} }
  & \; = \; &
  \frac{\sqrt{\lambda}}{\pi s^2}
  \left( {\cal C}^t {\cal G}^t \, + \,
         \sum_k {\cal C}^{st}_k {\cal G}^{st}_k\ \, + \,
         \sum_{ij} {\cal C}^s_{ij} {\cal  G}^s_{ij}
  \right)~~,
\label{anodiffxs}
\ea
where $i,j$, and $k$ run over $s,x,y,z$.
The summation over s-channel photon and $Z$ amplitudes is contained in
the coefficient functions ${\cal C}$.
Compared to the \sm\ where only three kinematical functions describe the
{\tt CC03} process, the inclusion of anomalous couplings adds
considerable complexity to the calculation.
The user may select a treatment of anomalous couplings with flag {\tt
  IANO}. 
The potential influence of an extra heavy neutral boson $Z'$ is
explained in appendix~\ref{extra}.
%
%
\section{Description of the Program}
\label{descr}
\ezero
{\tt GENTLE/4fan} is a {\sc Fortran77} program for the computation of
total and differential \xsecs\ for four-fermion production in \ee\
annihilation with inclusion of the relevant initial state QED
corrections. It originated from older versions~\cite{oldgen}.
%
\subsection{Features of ~{\tt GENTLE/4fan}}
The \xsecs, angular distributions, and moments presented in
equations~(\ref{sigres}), (\ref{totsig}), (\ref{ISRxstot}),
(\ref{strucxs}), (\ref{invmoments}), (\ref{angmoments}),
(\ref{angmomqed}), (\ref{angdiffqed}), and (\ref{anodiffxs}) are
computed in 
{\tt GENTLE/4fan} by numerically integrating analytical formulae for
expressions such as equation~(\ref{ISRxsdif}).
The analytical input formulae are found
in~
\cite{teupitz94,cc11,muta,nc24h,nc24,%
nuniall,nunicc,coulombdb,coulombdur,biebel}
and in references therein.
Numerical integrations are performed by a fast and stable
self-adaptive Simpson algorithm realized in the subroutines {\tt
  SIMPS} and 
{\tt FDSIMP}\footnote[5]{The {\sc Fortran} code {\tt FDSIMP} by Yu.~Sedykh
is an extension of the {\tt SIMPS} code by I.N~Silin.}.

For transparency -- and also for historical reasons -- version 2.0 of
{\tt GENTLE/4fan} is built from three pieces of code.
The corresponding ramification is controlled by the flag {\tt IPROC}. 
The first part contains the charged current ({\tt CC}) reactions
including \oal\ initial state QED corrections and anomalous couplings.
The second part is the {\tt 4fan} code with {\tt NC} reactions.
The third part, steered by the subroutine {\tt NCGENT} contains the
code for the {\tt NC02} and {\tt NC08} processes with
complete \oal\ initial state QED corrections.
Subsequently we will call the first part the {\tt CC} part, the second
one will be named the {\tt 4fan} branch, and the third part will be
called {\tt NCqed} part.
The first and second part together are referred to as the {\tt CC/NC}
branch.

At present, the {\tt GENTLE/4fan} may treat 
the following four-fermion final
states:
\begin{enumerate}
  \item[(1)] {\tt CC03} (with complete ISR, angular distributions, and
    anomalous 
    couplings)~\cite{nuniall,nunicc,biebel}
  \item[(2)] {\tt NC02}, {\tt NC08} (with complete
    ISR)~\cite{nuniall}
  \item[(3)] {\tt CC09}, {\tt CC10}, {\tt CC11}~\cite{cc11}
  \item[(4)] {\tt NC06}, {\tt NC10}, {\tt NC24}, {\tt NC32}~\cite{nc24}
  \item[(5)] {\tt NC} of (4) + Higgs~\cite{nc24h}
\end{enumerate}

Initial state radiation is implemented.
For total \xsecs\, {\em universal} initial state radiative corrections
as described in equations~(\ref{ISRxstot}) to~(\ref{softhard}) are
computed. 
In addition, {\em non-universal} ISR is available for the {\tt CC03},
{\tt NC02}, and {\tt NC08} processes.
For the {\tt CC11} family, leading logarithmic initial state QED
corrections can 
also be calculated in the structure function approach as indicated in
eq.~(\ref{angmomqed}).
Further, the Coulomb correction to {\tt CC03} and inclusive QCD
corrections (to both the denominators of the Breit-Wigner functions
and final states with quarks) have been implemented.
The user can choose between the different prescriptions of
references~\cite{coulombdb,coulombdur} for the Coulomb correction via
the flag {\tt ICOLMB}. 

Cuts may be applied on the invariant masses \sone\ and
\stwo\.
In the structure function approach, cuts on the electron and positron
momentum fractions after initial state radiation may be imposed.
In the {\tt NCqed} branch of the code, a minimum value for the fraction
${s'/s}$ may be set as well as for ${s_1/s}$ and ${s_2/s}$.
A cut on the {\tt CC03} $W$ production angle is not directly supported
by the code but may be imposed using the option for bin-integration.

In general, final state masses are neglected in {\tt GENTLE/4fan},
i.e. the program uses the small mass approximation.
Where needed, however, masses are retained in the phase space factors. 
Masses of heavy particles coupled to the Higgs boson are taken into
account where appropriate.
In the {\tt NCqed} branch of the program, masses may be chosen by
altering the values of {\tt AM1} or {\tt AM2} in the initialization
routine {\tt NCIN00}.
%
%
%
\subsection{Program input}
\label{input}
The common units throughout the code are GeV and GeV/$c^2$.
The physical input parameters of {\tt GENTLE/4fan} are defined in
subroutine {\tt WUFLAG}: \\
\begin{tabular}{rcccl}
     {\tt ALFAI} & = &  $\alpha(0)$  & = & 137.0359895, the
     inverse running fine structure \\
        & & & & constant at zero momentum transfer \\
     {\tt ALFAS}& = &  $\alpha_{_S}({\rm low}\: q^2)$ & = & $0.2$ \\
     {\tt ALPHS}& = &  $\alpha_{_S}(2M_W)$ & = & $0.12$ \\
     {\tt ALPW} & = &  $\alpha(2 M_W)$ & = &
        1/128.07, the running fine structure con- \\
        & & & & stant at $2 M_W$ \\
     {\tt AME}  & = &  $m_e$ & = &
        0.51099906 $\times 10^{-3}$ GeV, the electron mass \\
     {\tt AMZ}  & = &  $M_Z$ & = & 91.1888 GeV, the $Z$ mass \\
     {\tt AMW}  & = &  $M_W$ & = & 80.230 GeV, the $W$ mass \\
     {\tt GAMZ} & = &  $\Gamma_Z$ & = & 2.4974 GeV, the $Z$ width \\
     {\tt GFER} & = &  $G_\mu$ & = &
        1.16639 $\times 10^{-5}$ GeV$^{-2}$, the Fermi constant 
     \footnotemark[6]
\end{tabular}
\footnotetext[6]{ $\alpha_{_S}(2M_W)$\ is used for the QCD corrections to
                the $W$\ width and to the final state quark-anti-quark
                pairs in {\tt CC} processes.
                The low momentum transfer value $\alpha_{_S}(|q| \ll
                2M_W)$\ is used for the {\tt NC32}
                process~\cite{alteu96} in the calculation of gluon
                exchange diagrams.
 } 

For the fermion-boson couplings, the value $\alpha(2 M_W)$\ is used,
while for the ISR corrections, the Thompson limit $\alpha(0)$\ of the
fine structure constant is applied.
Depending on the flag value {\tt IINPT}, the weak mixing angle is
computed as
\ba
\nl
    {\tt SIN2W}  =  \sin^2 \theta_{W}  =&  
    {\displaystyle \frac{\pi \alpha(2 M_W)}{\sqrt{2} M_W^2 G_\mu}}
    & 
    {\rm for}~{\tt IINPT}=0~~
    \label{sw0}
\\
\nl 
    {\tt SIN2W} = \sin^2 \theta_{W}  =&  1 - M_W^2/M_Z^2
    & 
    {\rm for}~{\tt IINPT}=1~~
\\
\nl 
    {\tt SIN2W}  =  \sin^2 \theta_{W}  =& 0.22591
~{\rm (numerical~ user~ input)}  
    & 
    {\rm for}~{\tt IINPT}=2~~
    \label{sw1}
\ea

For charged current reactions, the following derived quantities are used
in {\tt GENTLE/4fan}:
\ba
  {\tt GAMW}  =  \Gamma_W = & 
    \frac{\displaystyle 9}{\displaystyle 6\sqrt{2}\pi} \,
    G_{\mu}M_W^3 
    \left( 1+\frac{\displaystyle 2 \alpha_{_S}(2M_W) }{\displaystyle
      3\pi} \right) & {\rm{for~~ }} {\tt IGAMW}=0
\label{gamw0}
    \\ \nl
  {\tt GAMW}  =  \Gamma_W = & 2.085~ 
    {\rm (numerical~ user~ input)~}  & {\rm for~ } {\tt IGAMW}=1
\label{gamw1}
\ea

as well as 

\begin{tabular}{rcccl}
  {\tt GAE} &  =  &$ - \frac{\displaystyle e}{\displaystyle 4s_Wc_W}$
  & = & 
    $- \frac{\displaystyle \sqrt{4\pi\alpha(2M_W)}}{\displaystyle
      4s_Wc_W}$ 
    \nl\nl
  {\tt GVE} &  =  & {\tt GAE}$   (1-4s_W)$& & 
    \nl\nl
  {\tt GWF} &  =  & $\frac{\displaystyle g}{\displaystyle 2\sqrt{2}}$
  & =& $-${\tt GAE}$  
  \sqrt{2} c_W $
    \nl\nl
  {\tt GWWG} & =  &$ \sqrt{4\pi\alpha(2M_W)}$&&
    \nl\nl
  {\tt GWWZ} &  =  & {\tt GWWG} $  {\displaystyle\frac{c_W}{s_W}}$ ~.
    \nonumber
\end{tabular}
\\

{\tt GVE} and {\tt GAE} are the electron vector and axial vector
couplings, {\tt GWF} is the fermion-$W$ coupling, and {\tt GWWG} and
{\tt GWWZ} are the triple gauge boson couplings for the photon
and the $Z$ boson in the Standard Model, respectively.
All other couplings are uniquely determined by the above.
Although, for historical reasons, some of the variable names are
slightly different, the same derived quantities are used for neutral
current reactions.
For the neutral current processes, the final state fermion types and
their masses are part of the input.
For the {\tt NC08} family of processes, this input is set in the
subroutine {\tt NCIN00}.
The fermion types for the {\tt NC32} family are set by the flags
{\tt IFERM1} and {\tt IFERM2}, while the corresponding masses are
loaded from a {\tt DATA} statement in the subroutine {\tt
  bbmmin}\footnote[7]{Note that the 
fermion masses are stored in the variables {\tt AM1} and {\tt AM2} for
the {\tt NC08} as well as the {\tt NC32} families. This is, however,
unproblematic, because the routines for each family are completely
separate and are never executed together in a single run.}.
In the subroutine {\tt WUFLAG}, the Higgs mass {\tt AMHIG} and the Higgs
width {\tt GAMHIG} as well as the flags which steer the program flow are
initialized.
The Higgs contribution may be switched on or off by setting 
flag {\tt IHIGGS}.


The flags and, for the {\tt CC} branch, the physical input may be
overridden by calls to the subroutine {\tt WUFLAG}; see also
section~\ref{use}.
{\em We strongly recommend the user to carefully control that only options
are active which are definitely declared to be compatible with each
other.}
\\ 
In {\tt WUFLAG}, the following flags are initialized:

\renewcommand{\arraystretch}{1.0}
\begin{tabular}{ll}
  \underline{{\tt IANO}} & \\
  \label{iano}
  {\tt IANO}:     & to set the anomalous couplings defined in
                    equations~(\ref{arrbeg}) \\
                  & to~(\ref{arrend}) of appendix~\ref{anocoup} \\
  {\tt IANO}=0:   & all couplings at their \sm\ value \\
  {\tt IANO}=$\pm$1: & {\tt XG} $= x_\gamma = \pm 0.5\,$,
                       all other couplings at their \sm\ values \\
  {\tt IANO}=$\pm$2: & {\tt YG} $= y_\gamma = \pm 0.5\,$,
                       all other couplings at their \sm\ values \\
  {\tt IANO}=$\pm$3: & {\tt XZ} $= x_Z = \pm 0.5\,$,
                       all other couplings at their \sm\ values \\
  {\tt IANO}=$\pm$4: & {\tt YZ} $= y_Z = \pm 0.5\,$,
                       all other couplings at their \sm\ values \\
  {\tt IANO}=$\pm$5: & {\tt ZZ} $= z_Z = \pm 0.1\,$,
                       all other couplings at their \sm\ values \\
  {\tt IANO}=$\pm$6: & {\tt DZ} $= \delta_Z = \pm 0.5\,$,
                       all other couplings at their \sm\ values \\
  {\tt IANO}=$\pm$7: & ``HISZ scenario''~\cite{hagiwara} with \\
                     & $ x_\gamma= \pm 0.1\,$, \\
                     & $x_Z = x_\gamma
                       \left(\cos^2\theta_W-\sin^2\theta_W\right)/
                       \left(2 \sin\!\theta_W \cos\!\theta_W\right)\,$, \\
                     & $\delta_Z =
                         x_\gamma/(2 \sin\!\theta_W \cos\!\theta_W)\,$, \\
                     & and $y_\gamma = y_Z = z_Z =0$ \\
  {\tt IANO}=8:    & user-set values of $x_\gamma$, $y_\gamma$, $x_Z$,
                     $y_Z$, $z_Z$, $\delta_\gamma$, and $\delta_Z$; \\
                   & for extra $Z'$ couplings, see appendix~\ref{extra}
  \vspace{.3cm} \\
  \underline{{\tt IBCKGR}} & \\
  {\tt IBCKGR}:   & only active for {\tt IPROC}=1 \\
  {\tt IBCKGR}=0: & calculations for the double resonant {\tt CC03}
                    process only, no \\
                  & {\tt CC11} background \\
  {\tt IBCKGR}=1: & inclusion of {\tt CC11} background in accordance
                    with {\tt ICHNNL} \\
                  & not available for {\tt IDCS}=1
  \vspace{.3cm}\\
  \underline{{\tt IBIN}} \\
  {\tt IBIN}:     & only active for {\tt IDCS}=1 \\
  {\tt IBIN}=0:   & calculation of $\d\sigma/\!\d\!\cos\!\theta$ for
                    {\tt IBINNU} fixed values of $\cos\!\theta$ \\
  {\tt IBIN}=1:   & calculation of $\d\sigma/\!\d\!\cos\!\theta$ integrated
                    over each of the \\
                  &  {\tt IBINNU} $\cos\!\theta$-bins
  \vspace{.3cm} \\
  \underline{{\tt IBINNU}} \\
  {\tt IBINNU}:   & only active for {\tt IDCS}=1 \\
  {\tt IBINNU}=n: & number of points or bins for the differential \xsec\
                     \\
                  & calculation
  \vspace{.3cm} \\
  \underline{{\tt IBORNF}} & \\
  {\tt IBORNF}=0: & Born approximation, no radiative corrections \\
  {\tt IBORNF}=1: & initial state QED corrections are included
  \vspace{.3cm}\\
\end{tabular}

\begin{tabular}{ll}
  \underline{{\tt ICHNNL}} & \\
  {\tt ICHNNL}:   & only active for {\tt IPROC}=1 and {\tt IBCKGR}=1 \\
  {\tt ICHNNL}=0: & no background taken into account, calculation of
                    inclusive \\
                  & {\tt CC03} quantities \\
  {\tt ICHNNL}=1: & calculation of {\tt CC09} quantities for a leptonic
                    final state \\
  {\tt ICHNNL}=2: & calculation of {\tt CC10} quantities for a
                    semi-leptonic final \\
                  & state with positively charged lepton \\
  {\tt ICHNNL}=3: & calculation of {\tt CC10} quantities for a
                    semi-leptonic final \\
                  & state with negatively charged lepton \\
  {\tt ICHNNL}=4: & calculation of {\tt CC11} quantities for a hadronic
                    final state \\
  {\tt ICHNNL}=5: & calculation of inclusive quantities for the {\tt
    CC11} family
  \vspace{.3cm} \\
  \underline{{\tt ICOLMB}} & \\
  {\tt ICOLMB}:   & determines, how the Coulomb correction is included in \\
                  & the differential \xsecs\ of equations~(\ref{diffsig})
                    or~(\ref{ISRxsdif}) \\
                  & only relevant for the {\tt CC03} process \\
  {\tt ICOLMB}=0: & Coulomb correction not included \\
  {\tt ICOLMB}=1: & Coulomb correction as in equation~(5) of
                    reference~\cite{coulombdb}, \\
                  & but without the phase factor \\
  {\tt ICOLMB}=2: & Coulomb correction as in equation~(5) of
                    reference~\cite{coulombdb},  \\
                  & but with $\Delta \equiv (\SONE-\STWO)/s' = 0$ \\
  {\tt ICOLMB}=3: & Coulomb correction as in equation~(5) of
                    reference~\cite{coulombdb} \\
  {\tt ICOLMB}=4: & Coulomb correction as in reference~\cite{coulombdurold} \\
  {\tt ICOLMB}=5: & Coulomb correction as in reference~\cite{coulombdur}
  \vspace{.3cm} \\
  \underline{{\tt ICONVL}} & \\
  {\tt ICONVL}:   & only active for {\tt IPROC}=1 and {\tt IPROC}=2
                    \\
  {\tt ICONVL}=0: & flux function convolution as given in
                    equation~(\ref{ISRxstot}) \\
  {\tt ICONVL}=1: & structure function convolution as indicated in
                    eq.~(\ref{strucxs})\\
                  & attention: extremely CPU time consuming
  \vspace{.3cm}\\
  \underline{{\tt IDCS}} & only active for {\tt IPROC}=1 and {\tt
                             IBCKGR}=0 \\ 
  {\tt IDCS}=0:   & calculation of total \xsecs\ and moments \\
  {\tt IDCS}=1:   & calculation of the differential \xsec
                    $\d\sigma/\!\d\!\cos\!\theta$ with \\
                  & anomalous couplings as determined by the flag {\tt IANO} \\
                  & if used in conjunction with ISR corrections, the 
                    structure \\
                  & function approach {\tt ICONVL}=1 must be used \\
                  & needs {\tt IBCKGR}=0 
  \vspace{.3cm} \\
\end{tabular}

\begin{tabular}{ll}
  \underline{{\tt IFERM1,$\,$IFERM2}} &  \\
  {\tt IFERM1}:      & only active for {\tt IPROC}=2 
                    {\tt IFERM1} is an integer specifying 
                     \\  
                  & 
                    the first final state fermion  
                    pair according to \\            
                  &  the PDG particle numbering
                    scheme~\cite{pdg96} \\
 {\tt IFERM2}:    & like {\tt IFERM1}, but for the
                        second final state  
                    fermion pair \\
                  & remember that it is the second final state fermion pair
                     \\
                  & that couples to the Higgs boson, if a calculation with the
                     \\
                  & Higgs is requested
  \vspace{.3cm} \\
  \underline{{\tt IGAMW}} \\
  {\tt IGAMW}=0:  & computation of the $W$ width according
                     to~(\ref{gamw0})
                  \\
  {\tt IGAMW}=1:  & numerical user input 
  \vspace{.3cm} \\
  \underline{{\tt IGAMZS}} & \\
  {\tt IGAMZS}=0: & constant $Z$ width for $Z$ propagators \\
  {\tt IGAMZS}=1: & $s$-dependent $Z$ width for  $Z$
                    propagators
  \vspace{.3cm} \\
  \underline{{\tt IHIGGS}} &  \\
  {\tt IHIGGS}:    & only active for {\tt IPROC}=2 \\
  {\tt IHIGGS}=0:  & Higgs not included\\
  {\tt IHIGGS}=1:  & Higgs included
 \vspace{.3cm} \\
   \underline{{\tt IINPT}} & \\  
  {\tt IINPT}=0:  & weak mixing angle defined by equation~(\ref{sw0}) \\
  {\tt IINPT}=1:  & weak mixing angle defined by equation~(\ref{sw1}) \\ 
  {\tt IINPT}=2:  & numerical user input, only active for {\tt IPROC}=1
  \vspace{.3cm}\\
  \underline{{\tt IIQCD}} &  \\
  {\tt IIQCD}:    & only active for {\tt IPROC}=1 \\
  {\tt IIQCD}=0:  & no QCD corrections, i.e. $\alpha_{_S}(2M_W) \equiv 0$ \\
  {\tt IIQCD}=1:  &  
                   inclusive QCD corrections to the $W$ width and
                    final state \\
                  & QCD corrections to the final state quark pair 
                    are taken \\
                  & into account with $\alpha_{_S}(2M_W) = 0.12$
  \vspace{.3cm} \\
  \underline{{\tt IMOMN}} \\
  {\tt IMOMN}:   & only active for {\tt IPROC}=1 and {\tt IDCS}=0 
                   \\
  {\tt IMOMN}=$n$: & the $n^{th}$\ moment of the physical quantity
                     determined by  \\
                 & the current {\tt IREGIM} is computed
\vspace{.3cm} \\
\end{tabular}
The moments to be computed are determined by the values of {\tt IMMIN}
and {\tt IMMAX} which control the scope of the {\sc Fortran} loop
  {\tt DO IMOMN=IMMIN,IMMAX}~. \vspace{.3cm}

\begin{tabular}{ll}
  \underline{{\tt INCPRC}} & \\
  {\tt INCPRC}:   & only active for {\tt IPROC}=3 \\
  {\tt INCPRC}=0: & {\tt NC02} \xsec\ calculation, i.e. $Z$ pair
  production 
                    only \\
  {\tt INCPRC}=1: & photon pair production only \\
  {\tt INCPRC}=2: & {\tt NC08} \xsec\ calculation
  \vspace{.3cm}\\
  \underline{{\tt IONSHL}} & \\
  {\tt IONSHL}:   & only active for {\tt IPROC}=1 \\
  {\tt IONSHL}=0: & on-shell limit for $W$ pair production \\
  {\tt IONSHL}=1: & off-shell $W$ bosons
  \vspace{.3cm}\\
  \underline{{\tt IPROC}} & \\
  {\tt IPROC}=1:  & {\tt CC11} family of processes \\
  {\tt IPROC}=2:  & {\tt NC32} family of processes \\
  {\tt IPROC}=3:  & {\tt NC08} family of processes with complete
                    ISR
  \vspace{.3cm}\\
  \underline{{\tt IQEDHS}} & \\
  {\tt IQEDHS}:   & determines the \SV\ and hard radiators ${\bar S}$\ and
                    ${\bar H}$\ to \\
                  & be used for universal corrections to total \xsecs\ \\
                  & only active, if {\tt IBORNF}=1 and {\tt ICONVL}=0 \\
  {\tt IQEDHS}=0: & \oal\ exponentiated universal ISR corrections as given \\
                  & in equation~(\ref{ISRxsdif}) with the radiators 
                    ${\bar S}$\ and ${\bar H}$\ from \\
                  & equations~(\ref{svrad}) and~(\ref{hardrad}) \\
  {\tt IQEDHS}=1: & as {\tt IQEDHS}=0, but terms of
                    ${\cal O}\left[ (\alpha L)^2\right]$ added to
                    ${\bar S}$\ and ${\bar H}$ \\
  {\tt IQEDHS}=2: & as {\tt IQEDHS}=1, but terms of
                    ${\cal O}\left( \alpha^2 L\right)$ added to
                    ${\bar S}$\ and ${\bar H}$ \\
  {\tt IQEDHS}=3: & as {\tt IQEDHS}=2, but terms of
                    ${\cal O}\left( \alpha^2 \right)$ added to
                    ${\bar S}$\ and ${\bar H}$ 
  \vspace{.3cm}\\
  \underline{{\tt IREGIM}} \\ 
\label{iremom}
  {\tt IREGIM}:   & only active for {\tt IPROC}=1 and {\tt IDCS}=0
                    \\
                  & determines the physical quantity to be computed in
                    addition \\
                  & to the total \xsec\ (see sections~\ref{qed}
                    and~\ref{angular}) \\
  {\tt IREGIM}=0: & computation of the total \xsec, \\
                  & used as normalization for next cases\\
  {\tt IREGIM}=1: & computation of moments of the radiative invariant mass
                    loss \\
  {\tt IREGIM}=2: & computation of moments of the radiative energy loss \\
                  & for the structure function approach, i.e. for
                    {\tt ICONVL}=1 only \\
  {\tt IREGIM}=3: & computation of moments of the $W$ mass shift for the 
                    {\tt CC03} \\
                  & process \\
  {\tt IREGIM}=4: & computation of $W$ production angular moments for the
                   {\tt CC03} \\
                  & process with Chebyshev polynomials (compare
                    section~\ref{angular})
\end{tabular}

The physical quantities to be computed are determined by the integer
values of the variables {\tt IRMAX} and {\tt IRSTP} which control the
scope of {\tt IREGIM} via the {\sc Fortran} statement
  {\tt DO IREGIM=0,IRMAX,IRSTP}. 

\vspace{.3cm}

\begin{tabular}{ll}
  \underline{{\tt ITNONU}} & \\
  {\tt ITNONU}:   & only active for {\tt IBORNF}=1 \\
  {\tt ITNONU}=0: & only universal hard ISR corrections are taken into \\
                  & account (compare equations~(\ref{softhard})
                    and~(\ref{hardrad}) ) \\
  {\tt ITNONU}=1: & complete universal and non-universal hard ISR cor-\\
                  & rections are taken into account;  only in effect for
                    \\
                  & {\tt CC03} and for the
                    {\tt NC08} family, {\tt IBCKGR}={\tt IDCS} = 0 \\
                  & attention: very CPU time consuming
  \vspace{.3cm}\\
  \underline{{\tt IZERO}} & \\
  {\tt IZERO}:    & only for the flux function convolution, i.e. for
                    {\tt ICONVL}=0 \\
  {\tt IZERO}=0:  & the constant term 
                    $(\alpha/\pi)(\pi^2/3 - 1/2 )$\ in the
                    ${\bar S}$\ is neglected  \\
                  & (compare equation~(\ref{svrad}) ) \\
  {\tt IZERO}=1:  & the \SV\ ${\bar S}$\ radiator is given by
                    equation~(\ref{svrad})
  \vspace{.3cm}\\
  \underline{{\tt IZETTA}} & \\
  {\tt IZETTA}:   & only for the structure function convolution, i.e. for
                    {\tt ICONVL}=1 \\
                  & compare equation~(4.21) of reference~\cite{cc11} \\
  {\tt IZETTA}=0: & {\tt ETA} choice, i.e. in the hard photon radiation
                    part of the \\
                  & structure function 
                    $(2\alpha/\pi) L$ is used,
                    whereas $(2\alpha/\pi) (L-1)$ is  \\
                  & entered into the soft part and the exponentiation \\
  {\tt IZETTA}=1: & {\tt BETA} choice, i.e. $(2\alpha/\pi) (L-1)$
                    is used for the whole struc\-tu\-re \\
                  & function
  \vspace{.3cm} \\
\end{tabular}

Finally, we mention that the relative and absolute precision
requirements for the numerical integrations are set at different
places in {\tt GENTLE/4fan}.
Sensible default values for the precision requirements are chosen for the
different branches of the program.
In addition, we collect here the physical input parameters which can
be chosen by calls to subroutine {\tt WUFLAG}:

\bigskip

{\tt ALPHS}, {\tt ALPW}, {\tt AME}, {\tt AMFER2}, {\tt AMHIG}, {\tt
  AMW}, {\tt AMZ}, {\tt DZ}, {\tt GAMHIG}, {\tt GAMW}, {\tt GAMZ},
{\tt GFER}, {\tt SINW2}, {\tt XG}, {\tt XZ}, {\tt YG}, {\tt YZ}, {\tt ZZ}.

\vspace{.5cm}
%
%
\subsection{Routines of {\tt GENTLE/4fan}}
\label{prgdesc}
In this subsection we give a brief description of the subroutines and
functions in the code.
{\tt GENTLE/4fan} uses {\tt REAL*8} and {\tt COMPLEX*16} variables.
\begin{itemize}
\itm {\tt FUNCTION alamk} -- computes the kinematic $\lambda$-function
     $\lambda(x,y,z)=x^2+y^2+z^2-2xy-2xz-2yz$
\itm {\tt FUNCTION ALRUN} -- may be used in subroutine {\tt prop} to
compute 
     photon propagators with running fine structure
     constant $\alpha$
\itm {\tt SUBROUTINE bbmmin} -- constitutes the central routine for the
     {\tt 4fan} part of the code, i.e. for computations for the
     {\tt NC32} family; performs the initialization of physical and
     coupling constants for the {\tt 4fan} part
\itm {\tt FUNCTION c222 (c322, c322m, c422, c233)} ~--~ computes the
     coupling function ${\cal C}_{222}$ (${\cal C}_{322}$,
     ${\cal C}^a_{322}$, ${\cal C}_{422}$, ${\cal C}_{233}$) introduced
     in references~\cite{nc24h,nc24}\footnote[8]{Beware: The
     normalization differs from the references in some cases.}
\itm {\tt FUNCTION cg422 (cg233)} -- computes the coupling function
     similar to ${\cal C}_{422}$ (${\cal C}_{233}$), but with a gluon
     exchange replacing the electroweak boson exchange in the final\
     state (compare figure~1e of reference~\cite{alteu96})
\itm {\tt SUBROUTINE check} -- checks and prints the input to the
     {\tt 4fan} part; produces the integer error flag {\tt ierr};
     prints messages if input errors are encountered  
\itm {\tt FUNCTION COULOM} -- computes, according to the flag value of
     {\tt ICOLMB}, the term {\tt COULOM} for the Coulomb correction
     factor $\left(1\!+\!\right.${\tt COULOM}$\left.\right)$
\itm {\tt FUNCTIONS DDILOG, DLI2} -- provide the dilogarithm function
     Li$_2(x)$ for real argument; if $x \geq 1$, $x$ is understood to
     be the real part of the argument with infinitesimal imaginary
     part, and the real part of the function value is returned
\itm {\tt SUBROUTINE DELTPR} -- computes, for the flux function
     approach, the threefold (with respect to \sone, \stwo, and $s'$)
     differential, non-universal contributions to the {\tt CC03} \xsec
\itm {\tt SUBROUTINE DELVNU} -- computes the non-universal \SV\
     contribution to the {\tt NCqed} \xsec
\itm {\tt FUNCTION DLI3} -- provides the trilogarithm function
     Li$_3(x)$ for $x \leq 1$
\itm {\tt FUNCTION DS12} -- provides the Nielsen function
     S$_{1,2}(x)$ for $x \leq 1$
\itm {\tt SUBROUTINE dsdshsz} -- provides the twofold differential
     \xsec\ 
      \\
      $\d^2 \sigma / (\d\SONE \d\STWO) $ for the {\tt NC32} family
     as given in references~\cite{nc24h,nc24}
\itm {\tt FUNCTION DXS2} -- provides, for the {\tt CC/NC} branch of the
     code and according to the flags {\tt IONSHL}, {\tt IPROC},
     {\tt IBORNF}, {\tt IBCKGR}, and {\tt ICHNNL}, twofold differential
     expressions  
     (with respect to \sone\ and \stwo) for the \xsec\ or for moments
     by calling {\tt XSECB} and {\tt XSECBG} or {\tt XSEC2E}
\itm {\tt FUNCTION FACINC} -- is the inverse function of {\tt FACNC}
\itm {\tt FUNCTION FACINV} -- represents the inverse function of
     {\tt FACT}
\itm {\tt FUNCTION FACNC} -- is the mapping function for the $s'$
     integrations used in calls to {\tt FDSIMT}
\itm {\tt FUNCTION FACT} -- is the mapping function for the $s'$,
     $x_+$, and $x_-$ integrations used in calls to {\tt FDSIMP} and
     {\tt FDSIMS}
\itm {\tt FUNCTION FCONC} -- provides the integrand of the $s'$
     integration in {\tt XSNC2E}
\itm {\tt FUNCTION FCONV1} -- provides, according to the flag
     {\tt ICONVL}, threefold differential (with respect to \sone, \stwo,
     and $s'$\ or $x_+$) expressions for
     \begin{quotation}
     \begin{enumerate}
       \item the ISR-corrected \xsec, the radiative invariant mass loss,
             and the $W$ mass shift using the flux function approach
       \item total and differential \xsecs\ as well as moments in the
             structure function approach by calling {\tt FCONV2}
     \end{enumerate}
     \end{quotation}
\itm {\tt FUNCTION FCONV2} -- provides fourfold differential (with
     respect to \sone, \stwo, $x_+$, and $x_-$) expressions for all
     quantities available in the structure function approach
\itm {\tt SUBROUTINE FDSIMP} ~--~ is
     similar to {\tt SIMPS}, but performs the integration with the help
     a mapping function which is called in the argument list
\itm {\tt SUBROUTINES FEYNINT1, FEYNINT2, FEYNINT3} -- compute, from the
     analytical formulae of reference~\cite{nuniall}, the values of
     several complicated integrals needed for the non-universal hard ISR
     correction
\itm {\tt FUNCTION FINE12 (FI12NC)} ~~--~~ provides the differential
     \xsec\ $\d \sigma/\d x_1$, $x_i = s_i/s$ for the {\tt CC/NC}
     ({\tt NCqed}) branch of the code
\itm {\tt FUNCTION FINE34 (FI34NC)} ~~--~~ provides the differential
     \xsec\ $\d \sigma/(\d x_1 \d x_2)$\ for the {\tt CC/NC}
     ({\tt NCqed}) branch of the code by calling the function {\tt DXS2}
     ({\tt XSBRNC} and {\tt XSNC2E}) which actually
     provides the value
\itm {\tt FUNCTION FMAPSH} -- performs, for the {\tt NC} branch, the
     mapping in the invariant mass squared $s_2$ of the fermion pair
     defined by the flag {\tt IFERM2}
\itm {\tt FUNCTION FMAPSZ} -- performs, for the {\tt NC} branch, the
     mapping in the invariant mass squared $s_1$ of the fermion pair
     defined by the flag {\tt IFERM1}
\itm {\tt FUNCTION FMBW} -- 
     maps Breit-Wigner resonance peaks
\itm {\tt FUNCTION FMQED} -- 
     maps peaks of the type $1/s$
\itm {\tt FUNCTION FOTF} -- computes the leptonic vacuum polarization for 
     {\tt ALRUN} from lepton and effective quark masses
\itm {\tt FUNCTION FOTF1} -- computes the leptonic vacuum polarization for 
     {\tt ALRUN} from lepton masses and the hadronic vacuum polarization
     of reference~\cite{hadpol}
\itm {\tt FUNCTION FSZ} -- performs, for the {\tt NC} branch, the
     integration over $s_2$; this is the complement to the function
     {\tt FINE12}
\itm {\tt FUNCTION FTBINV} -- represents the inverse function of
     {\tt FXTB}
\itm {\tt FUNCTION FXSECB} -- provides, in dependence on the flag
     {\tt IDCS} and for the {\tt CC03} process, the threefold
     differential \xsec\ $\d^3 \sigma/(\d x_1 \d x_2 \d \cos\!\theta)$\
     or the argument $T_n (\cos\!\theta_{lab})  
     \d^3\sigma_{\tt CC03}/(\d \SONE \d \STWO \d\cos\!\theta)$\
     of the moment integral~(\ref{angmomqed})
\itm {\tt FUNCTION FXSECD} -- provides, in dependence on the flags
     {\tt IDCS} and {\tt IBIN} and for the {\tt CC03} process, the
     integral over the 
     threefold differential \xsec\ $\d^3 \sigma/(\d x_1 \d x_2 \d
     \cos\!\theta)$\ for given limits     
\itm {\tt FUNCTION FXTB} -- is the mapping function for the
     $\cos\!\theta$ integration with the subroutine {\tt FDSIMV}
\itm {\tt FUNCTION g222 (g322, g322m, g422, g233)} -- computes the
     kine\-matic function ${\cal G}_{222}$ (${\cal G}_{322}$,
     ${\cal G}^a_{322}$, ${\cal G}_{422}$, ${\cal G}_{233}$) introduced
     in references~\cite{nc24h,nc24}
\itm {\tt SUBROUTINE GFBGD2} -- computes the kinematic functions
     ${\cal G}_k$ for the pure background for the {\tt CC11} process
\itm {\tt SUBROUTINE GFBGD3} -- computes the kinematic functions
     ${\cal G}_k$ for the signal-background interferences of the
     {\tt CC11} process
\itm {\tt SUBROUTINE GFBRN} -- computes the kinematic functions
     ${\cal G}_k$\ for the {\tt CC03} process (compare
     equation~(\ref{diffsig}), see reference~\cite{muta} )
\itm {\tt SUBROUTINE GFBRNA} -- computes, in dependence on the flags
     {\tt IDCS} and {\tt IANO}, the angular kinematic functions
     ${\cal G}^{\tt CC03}_k(s;\SONE,\STWO,\cos\!\theta)$\ (compare
     eq.~(\ref{sigstc}), see reference~\cite{cc11} )
\itm {\tt SUBROUTINE GFBRNB} -- computes, in dependence on the flags
     {\tt IDCS} and {\tt IANO}, the integrated angular kinematic functions
     ${\cal G}^{\tt CC03}_k(s;\SONE,\STWO,\cos\!\theta_{\rm
     min},\cos\!\theta_{\rm max})$\ 
\itm {\tt SUBROUTINE GFBRNC} -- computes the kinematic function for the
     {\tt NCqed} branch
\itm {\tt MAIN} -- performs the ramification between the {\tt NCqed} and
     the {\tt CC/NC} branches of the code; carries out some
     initialization; calls {\tt WWIN00} and {\tt WWIN01} for further
     initialization; loops over the center of mass energies; loops over
     the desired values of {\tt IREGIM} and {\tt IMOMN}; prints results
\itm {\tt FUNCTION MYEXP1 (MYEXP2)} -- is the inverse function of
     {\tt MYLOG1} ({\tt MYLOG2})
     as physics input
\itm {\tt FUNCTION MYLOG1 (MYLOG2)} -- is the mapping function for the
     \sone (\stwo) integration used in calls to {\tt FDSIMP}
     ({\tt FDSIMS})
\itm {\tt SUBROUTINE NCGENT} -- the central routine for the
     {\tt NCqed} branch of the code; sets integration precisions; calls
     {\tt NCIN00} and {\tt NCIN01} for further initialization; loops
     over the center of mass energies; provides the total \xsec\ for the
     {\tt NCqed} branch; prints results
\itm {\tt SUBROUTINE NCIN00} -- initializes the flags needed for the
     {\tt NCqed} branch of the code; initializes the fermion types and
     couplings and performs other kinematics independent initialization
     for the {\tt NCqed} branch
\itm {\tt SUBROUTINE NCIN01} -- performs, for the {\tt NCqed} branch of
     the code, the initialization of quantities that depend on $s$ only
     and provides the \SV\ radiator
\itm {\tt SUBROUTINE NCIN02} -- provides the phase space factor and the
     coupling function for the {\tt NCqed} branch given in
     equation~(4.5) of first of references~\cite{nuniall}
\itm {\tt SUBROUTINE prop} -- provides the $Z$, $\gamma$, and gluon
     propagators for use in the subroutine {\tt dsdshsz}
\itm {\tt SUBROUTINE PROPAG} -- computes the neutral gauge boson
     propagators used in {\tt NCIN02}
\itm {\tt FUNCTION RHOINV} -- is the inverse function of {\tt RHOPR}
\itm {\tt FUNCTION RHOPR} -- provides, for the {\tt CC/NC} branch of the
     code and according to the mapping choice via {\tt IMAPPG}, the
     mapping function for the \sone\ and \stwo\ integrations
\itm {\tt FUNCTION RI3} -- computes the scalar two-point function needed
     in {\tt FOTF} and {\tt FOTF1}
\itm {\tt SUBROUTINE SETANO} -- initializes the anomalous couplings
     defined by equations~(\ref{arrbeg}) to~(\ref{arrend}) in
     dependence on the flags {\tt IDCS} and {\tt IANO}
\itm {\tt SUBROUTINE SHORTI} -- computes the kinematic functions
     ${\cal G}^{uu,dd}_{\tt CC11}$\ and ${\cal G}^{u,d}_{\tt CC11}$\
     as given in equations~(3.9) and~(3.10) of reference~\cite{cc11}
\itm {\tt SUBROUTINE SIMPS, SIMPT} -- performs the integration of an input
     function inside given limits and with a given precision by applying
     a self-adaptive Simpson algorithm
\itm {\tt FUNCTION SPENC} -- is an auxiliary function used by the
     complex dilogarithm function {\tt XSPENZ}
\itm {\tt FUNCTION STRUCF} -- computes, according to the value of  
     {\tt IZETTA}, the structure function $D(x)$ from appendix A of
     reference~\cite{LEP2WW}
\itm {\tt FUNCTION TOTXS} -- computes the total \xsec\ for the
     {\tt CC/NC} branch and the differential \xsec\ for the
     {\tt CC03} case
\itm {\tt FUNCTION TRILOG} -- provides the trilogarithm function
     Li$_3(x)$ for $0 \leq x \leq 1$
\itm {\tt FUNCTION TRIS12} -- provides the Nielsen function
     S$_{1,2}(x)$ for $0 \leq x \leq 1$
\itm {\tt SUBROUTINE WUFLAG} -- initializes and changes the default
     flags as well as physics input
\itm {\tt SUBROUTINE WWIN00} -- performs the kinematics independent
     initialization of mathematical and physical constants for the
     {\tt CC/NC} branch of the code; calls the subroutines {\tt bbmmin}
     and {\tt SETANO} for further initialization
\itm {\tt SUBROUTINE WWIN01} -- performs, for the {\tt CC/NC} branch,
     the initialization of quantities that depend on the center of mass
     energy or, equivalently, $s$ only; performs initialization
     in dependence on the flags {\tt IPROC}, {\tt ICONVL}, and
     {\tt IQEDHS}; provides the \SV\ radiator for the {\tt CC/NC}
     computation
\itm {\tt SUBROUTINE WWIN02} -- performs the initialization of tree
     level quantities that depend on $s,~\SONE$, and $\STWO$; for the
     {\tt CC/NC} branch
\itm {\tt SUBROUTINE WWIN03} -- performs the initialization of radiative
     quantities that depend on $s,~\SONE$, and $\STWO$; for the
     {\tt CC/NC} branch
\itm {\tt SUBROUTINE XSBRNC} -- provides the tree level \xsec\ for the
     {\tt NCqed} branch of the code
\itm {\tt SUBROUTINE XSEC2E} -- is needed for the computation of 
     ISR-corrected \xsecs, moments, and distributions; performs the
     $s'$\ integration for the flux function approach and the $x_+$\
     integration for the structure function approach 
\itm {\tt SUBROUTINE XSEC3B} -- computes, in the flux function approach
     and for the {\tt CC11} family with {\tt IBCKGR}=1, the universally
     ISR-corrected, threefold (with respect to \sone, \stwo, and $s'$)
     differential \xsec\ for background contributions by calling the
     subroutine {\tt XSECBG}
\itm {\tt SUBROUTINE XSEC3E} ~--~ is used in the flux function
     convolution for the {\tt CC03} process, i.e. for {\tt IPROC=1},
     {\tt IBORNF}=1, and {\tt ICONVL}=0 only;
     computes, in dependence on the settings of 
     {\tt ITNONU} and {\tt IQEDHS} and according to
     formula~(\ref{ISRxsdif}) and reference~\cite{nunicc}, the threefold
     differential {\tt CC03}-expressions
     $\left[ \beta_e v^{\beta_e - 1} {\cal S}_k+{\cal H}_k \right]$\
\itm {\tt SUBROUTINE XSECB} -- for the {\tt CC/NC} branch; provides, in
     dependence on the flags {\tt IPROC}, {\tt IDCS}, and {\tt IREGIM},
     the twofold differential (with respect to \sone\ and \stwo) \xsec\
     or moments for the chosen tree level process by calling {\tt GFBRN}
     or integrating {\tt FXSECB} over the $W$ production angle 
\itm {\tt SUBROUTINE XSECBG} -- provides, in dependence on the channel
     chosen via the flag {\tt ICHNNL}, the twofold (with respect to
     \sone\ and \stwo) differential \xsec\ for the signal-background
     interference and the pure background in the {\tt CC11} process by
     calling the subroutines {\tt GFBGD2} and {\tt GFBGD3}
\itm {\tt SUBROUTINE XSNC2E} -- provides the twofold (with respect to
     \sone\ and \stwo) differential \xsec\ for the radiative
     {\tt NCqed} \xsec\ by integrating over $s'$
\itm {\tt SUBROUTINE XSNC3E} -- computes, according to the flags
     {\tt ITNONU}, the threefold (with respect to
     \sone\ and \stwo) differential, radiative {\tt NCqed} \xsec\
\itm {\tt FUNCTION XSPENZ} -- provides the complex dilogarithm function
     Li$_2(z)$ by transforming the argument to the area of fast
     convergence
\end{itemize}
%
\subsection{Program output}
\label{output}
The output of {\tt GENTLE/4fan} strongly depends on the user-chosen
branch of the program or, in other words, on the flag settings.
Four different output appearances are to be distinguished, namely
\begin{enumerate}
\item the output of total \xsecs\ and moments for the {\tt CC11} family
\item the output of angular distributions for the {\tt CC03} process
\item the output of the {\tt 4fan} branch of the code
\item the output of the {\tt NCqed} branch of the code
\end{enumerate}
In the below subsections, we will provide and explain examples for each
of the above four kinds of output.
The below outputs will, at the same time, serve as test run output, where
we assume that the user has used the flags and physical input as shown
in the output.
Any other pre-set input should remain unchanged for test run purposes.
\subsubsection{Output of total \xsecs\ and moments for {\tt CC11}}
The output of {\tt GENTLE/4fan} for the total \xsec\ and moments for
the {\tt CC11} family of processes appears as follows.
\begin{verbatim}
 ***** This is GENTLE/4fan -- Version 2.0 *****

 This is the CC branch of GENTLE/4fan for total cross-sections
 FLAGS:IPROC ,IINPT ,IONSHL,IBORNF,IBCKGR,ICHNNL= 1 1 1 1 1 2
 FLAGS:IGAMZS,IGAMW ,ITNONU,IQEDHS,ICOLMB,ICONVL= 1 0 0 1 2 0
 FLAGS:IZERO ,IZETTA,IIQCD ,IDCS  ,IANO  ,IBIN  = 0 1 1 0 0 0
 FLAGS:IMAP  ,IRMAX ,IRSTP ,IMMIN ,IMMAX        = 1 1 1 1 4
 EPS1 =  1.0000000000000000E-04
 S1MIN,S1MAX=  0.0000E+00  0.2592E+05
 S2MIN,S2MAX=  0.0000E+00  0.2592E+05
 Energy (GeV)  Cross-section (pb)
    161.00       0.1336661D+00
 MOMENTS
  1           1              2              3              4
  1   0.4753809D+00  0.3387719D+01  0.5607608D+02  0.1411261D+04
\end{verbatim}

\clearpage

\begin{verbatim}
 S1MIN,S1MAX=  0.0000E+00  0.3063E+05
 S2MIN,S2MAX=  0.0000E+00  0.3063E+05
 Energy (GeV)  Cross-section (pb)
    175.00       0.4952086D+00
 MOMENTS
  1           1              2              3              4
  1   0.1126753D+01  0.8432371D+01  0.1030576D+03  0.1885898D+04
 S1MIN,S1MAX=  0.0000E+00  0.3610E+05 
 S2MIN,S2MAX=  0.0000E+00  0.3610E+05
 Energy (GeV)  Cross-section (pb)
    190.00       0.6080011D+00
 MOMENTS
  1           1              2              3              4
  1   0.2151788D+01  0.2755849D+02  0.4972633D+03  0.1089393D+05
 S1MIN,S1MAX=  0.0000E+00  0.4203E+05
 S2MIN,S2MAX=  0.0000E+00  0.4203E+05
 Energy (GeV)  Cross-section (pb)
    205.00       0.6355688D+00
 MOMENTS
  1           1              2              3              4
  1   0.3208536D+01  0.5908307D+02  0.1485244D+04  0.4348380D+05
\end{verbatim}
\vspace{.5cm}
\begin{verbatim}
 ***** This is GENTLE/4fan -- Version 2.0 *****

 This is the CC branch of GENTLE/4fan for total cross-sections
 FLAGS:IPROC ,IINPT ,IONSHL,IBORNF,IBCKGR,ICHNNL= 1 1 1 1 0 2
 FLAGS:IGAMZS,IGAMW ,ITNONU,IQEDHS,ICOLMB,ICONVL= 1 0 0 1 2 0
 FLAGS:IZERO ,IZETTA,IIQCD ,IDCS  ,IANO  ,IBIN  = 0 1 1 0 0 0
 FLAGS:IMAP  ,IRMAX ,IRSTP ,IMMIN ,IMMAX        = 1 3 3 1 4
 EPS1 =  1.0000000000000000E-04
 S1MIN,S1MAX=  0.0000E+00  0.2592E+05
 S2MIN,S2MAX=  0.0000E+00  0.2592E+05
 Energy (GeV)  Cross-section (pb)
    161.00       0.1332829D+00
 MOMENTS
  3           1              2              3              4
  3  -0.3052332D+01  0.2605644D+02 -0.3833234D+03  0.7596405D+04
 S1MIN,S1MAX=  0.0000E+00  0.3063E+05
 S2MIN,S2MAX=  0.0000E+00  0.3063E+05
 Energy (GeV)  Cross-section (pb)
    175.00       0.4946375D+00
 MOMENTS
  3           1              2              3              4
  3  -0.5666559D+00  0.8922053D+01 -0.9553119D+02  0.1918868D+04
\end{verbatim}

\clearpage

\begin{verbatim}
 S1MIN,S1MAX=  0.0000E+00  0.3610E+05
 S2MIN,S2MAX=  0.0000E+00  0.3610E+05
 Energy (GeV)  Cross-section (pb)
    190.00       0.6074997D+00
 MOMENTS
  3           1              2              3              4
  3  -0.1004648D+00  0.9783632D+01 -0.4987223D+02  0.1646710D+04
 S1MIN,S1MAX=  0.0000E+00  0.4203E+05
 S2MIN,S2MAX=  0.0000E+00  0.4203E+05
 Energy (GeV)  Cross-section (pb)
    205.00       0.6352834D+00
 MOMENTS
  3           1              2              3              4
  3   0.1802810D+00  0.1232392D+02  0.6206403D+01  0.2242722D+04
\end{verbatim}

First, the program identifies itself and the branch of the code for
which output is produced.
Then, all relevant flags are printed, before the relative precision
{\tt EPS1} of the last computation is given.
Finally, physics output is generated.
For each center of mass energy the energy and the corresponding
total \xsec\ are printed.
Then, in the first column, integers $k$\ indicate the value of
{\tt IREGIM} and thus determine the physics quantity for which moments
are output.
The first line in the moments' output represents the degree $n$\ of the
moments in the column below.
In the second line, the moment of degree $n$\ for the quantity determined
by $k$\ is given.
Its value is normalized to the total \xsec.
\subsubsection{Output of differential \xsecs\ for {\tt CC03}}
First we give an example of output for {\tt IBIN}=0, with differential
\xsecs\ computed at fixed values for the scattering angle.
\begin{verbatim}
 ***** This is GENTLE/4fan -- Version 2.0 *****

 This is the CC branch of GENTLE/4fan for diff. cross-sections
  XG=    0.10000D+00   XZ=   -0.10000D+00
  YG=    0.00000D+00   YZ=    0.00000D+00
  ZZ=    0.00000D+00   DZ=    0.50000D-01
 FLAGS:IPROC ,IINPT ,IONSHL,IBORNF,IBCKGR,ICHNNL= 1 1 1 0 0 2
 FLAGS:IGAMZS,IGAMW ,ITNONU,IQEDHS,ICOLMB,ICONVL= 1 0 0 1 2 1
 FLAGS:IZERO ,IZETTA,IIQCD ,IDCS  ,IANO  ,IBIN  = 0 1 1 1 8 0
 FLAGS:IMAP  ,IRMAX ,IRSTP ,IMMIN ,IMMAX        = 1 3 5 1 1
 EPS1 =  1.0000000000000000E-04
\end{verbatim}

\clearpage

\begin{verbatim}
 S1MIN,S1MAX=  0.0000E+00  0.2592E+05
 S2MIN,S2MAX=  0.0000E+00  0.2592E+05
 Energy (GeV)=   161.0000000000000    
   COSW     Cross-section (pb)
  -1.000      0.0490713
  -0.800      0.0536632
  -0.400      0.0646671
   0.000      0.0794040
   0.400      0.1009377
   0.800      0.1381398
   1.000      0.1760354
 S1MIN,S1MAX=  0.0000E+00  0.3063E+05
 S2MIN,S2MAX=  0.0000E+00  0.3063E+05
 Energy (GeV)=   175.0000000000000    
   COSW     Cross-section (pb)
  -1.000      0.0951087
  -0.800      0.1123347
  -0.400      0.1564827
   0.000      0.2238638
   0.400      0.3411499
   0.800      0.5873924
   1.000      0.8353949
 S1MIN,S1MAX=  0.0000E+00  0.3610E+05
 S2MIN,S2MAX=  0.0000E+00  0.3610E+05
 Energy (GeV)=   190.0000000000000    
   COSW     Cross-section (pb)
  -1.000      0.0710393
  -0.800      0.0891843
  -0.400      0.1359378
   0.000      0.2124469
   0.400      0.3669140
   0.800      0.7949695
   1.000      1.3546129
 S1MIN,S1MAX=  0.0000E+00  0.4203E+05
 S2MIN,S2MAX=  0.0000E+00  0.4203E+05
 Energy (GeV)=   205.0000000000000    
   COSW     Cross-section (pb)
  -1.000      0.0512161
  -0.800      0.0678664
  -0.400      0.1099483
   0.000      0.1806793
   0.400      0.3368272
   0.800      0.8655771
   1.000      1.7575708
\end{verbatim}
After the identification of the active {\tt GENTLE/4fan} branch, the
anomalous couplings (compare page~\pageref{iano} and
appendix~\ref{anocoup}) used in the {\tt CC03} subprocess are output,
before the used flags and the relative precision of the run are printed.
Next, below the print of each center of mass energy, the differential
\xsecs\ for {\tt IBINNU} fixed values of the scattering angle {\tt COSW}
are written to the output.

Below, we present output for {\tt IBIN}=1, i.e the given differential 
\xsec\ values are integrated over bins in the scattering angle.
The appearance of the output is identical to the above output for
{\tt IBIN}=0 except that, for each energy, the total \xsec\ is given
right below the prints of scattering angle values and the corresponding
\xsecs.

\begin{verbatim}
 ***** This is GENTLE/4fan -- Version 2.0 *****

 This is the CC branch of GENTLE/4fan for diff. cross-sections
  XG=    0.10000D+00   XZ=    0.65543D-01
  YG=    0.00000D+00   YZ=    0.00000D+00
  ZZ=    0.00000D+00   DZ=    0.11957D+00
 FLAGS:IPROC ,IINPT ,IONSHL,IBORNF,IBCKGR,ICHNNL= 1 1 1 0 0 2
 FLAGS:IGAMZS,IGAMW ,ITNONU,IQEDHS,ICOLMB,ICONVL= 1 0 0 1 2 1
 FLAGS:IZERO ,IZETTA,IIQCD ,IDCS  ,IANO  ,IBIN  = 0 1 1 1 7 1
 FLAGS:IMAP  ,IRMAX ,IRSTP ,IMMIN ,IMMAX        = 1 3 5 1 1
 EPS1 =  1.0000000000000000E-04
 S1MIN,S1MAX=  0.0000E+00  0.2592E+05
 S2MIN,S2MAX=  0.0000E+00  0.2592E+05
 Energy (GeV)=   161.0000000000000    
       COSW        Cross-section (pb)
 -1.000 - -0.600    0.020923
 -0.600 - -0.200    0.025547
 -0.200 -  0.200    0.031763
  0.200 -  0.600    0.040919
  0.600 -  1.000    0.057338
    TOTAL XSEC =    0.176489
 S1MIN,S1MAX=  0.0000E+00  0.3063E+05
 S2MIN,S2MAX=  0.0000E+00  0.3063E+05
 Energy (GeV)=   175.0000000000000    
       COSW        Cross-section (pb)
 -1.000 - -0.600    0.042294
 -0.600 - -0.200    0.060777
 -0.200 -  0.200    0.089224
  0.200 -  0.600    0.139347
  0.600 -  1.000    0.247555
    TOTAL XSEC =    0.579198
\end{verbatim}

\clearpage

\begin{verbatim}
 S1MIN,S1MAX=  0.0000E+00  0.3610E+05
 S2MIN,S2MAX=  0.0000E+00  0.3610E+05
 Energy (GeV)=   190.0000000000000    
       COSW        Cross-section (pb)
 -1.000 - -0.600    0.032453
 -0.600 - -0.200    0.051912
 -0.200 -  0.200    0.084241
  0.200 -  0.600    0.150952
  0.600 -  1.000    0.346228
    TOTAL XSEC =    0.665786
 S1MIN,S1MAX=  0.0000E+00  0.4203E+05
 S2MIN,S2MAX=  0.0000E+00  0.4203E+05
 Energy (GeV)=   205.0000000000000    
       COSW        Cross-section (pb)
 -1.000 - -0.600    0.023983
 -0.600 - -0.200    0.041371
 -0.200 -  0.200    0.071246
  0.200 -  0.600    0.139243
  0.600 -  1.000    0.393478
    TOTAL XSEC =    0.669320
\end{verbatim}
\subsubsection{Output for the {\tt 4fan} branch}
\begin{verbatim}
 ***** This is GENTLE/4fan -- Version 2.0 *****

 This is the 4fan branch of GENTLE/4fan
 4fan was called for fermions 13  5 
 with masses   0.500000D-03  0.100000D-01
 total cross section is calculated 

 FLAGS:IPROC ,IINPT ,IONSHL,IBORNF,IBCKGR,ICHNNL= 2 0 1 0 0 0
 FLAGS:IGAMZS,IGAMW ,ITNONU,IQEDHS,ICOLMB,ICONVL= 1 0 0 1 0 0
 FLAGS:IZERO ,IZETTA,IIQCD ,IDCS  ,IANO  ,IBIN  = 1 1 0 0 0 0
 FLAGS:IMAP  ,IRMAX ,IRSTP ,IMMIN ,IMMAX        = 1 0 1 1 1
 FLAGS FOR 4FAN: IFERM1, IFERM2, IHIGGS=13 5 0
 EPS1 =  1.0000000000000000E-04
 ABS1 =  1.0000000000000000E-04
 S1MIN,S1MAX=  0.5805E+04  0.1128E+05
 S2MIN,S2MAX=  0.9000E+03  0.2592E+05
\end{verbatim}
\clearpage
\begin{verbatim}
 Energy (GeV)  Cross-section (pb)
    161.00       0.2907564D+00
 S1MIN,S1MAX=  0.5805E+04  0.1128E+05
 S2MIN,S2MAX=  0.9000E+03  0.3063E+05
 Energy (GeV)  Cross-section (pb)
    175.00       0.8156756D+00
 S1MIN,S1MAX=  0.5805E+04  0.1128E+05
 S2MIN,S2MAX=  0.9000E+03  0.3610E+05
 Energy (GeV)  Cross-section (pb)
    190.00       0.9510082D+01
 S1MIN,S1MAX=  0.5805E+04  0.1128E+05
 S2MIN,S2MAX=  0.9000E+03  0.4203E+05
 Energy (GeV)  Cross-section (pb)
    205.00       0.1249948D+02
\end{verbatim}

\subsubsection{Output for the {\tt NCqed} branch}
\begin{verbatim}
 ***** This is GENTLE/4fan -- Version 2.0 *****

 This is the NCqed branch of GENTLE/4fan
 FLAGS: NINPT,NNCPRC,NGAMZS,NBORNF,NTNONU,NQEDHS
          0      2      1      1      0      0

 GFER = 1.166390000000000E-05
 AMW  = 80.23
 AMZ  = 91.1888
 GAMZ = 2.4974
 ALPW = 7.808229874287499E-03
 SIN^2(\theta_W) = .2310309124515784

 IFERM1,IFERM2   =      2             4
 RNCOU1,RNCOU2   =  1.00000000    3.00000000
 AM1   ,AM2      = .105658389 4.3
 CUTM12,CUTM34   = .211316778 8.6
 CUTXPR = .0

 EPS1 1.000000000000000E-05
 ABS1 1.000000000000000E-15
 ENERGY (GeV)    CROSS-SECTION (fb)
    161.0          .8105861091D+02
    175.0          .6725488126D+02
    190.0          .6303702520D+02
    205.0          .5842607173D+02
\end{verbatim}
After identifying the active branch, the flags relevant
to the {\tt NCqed} branch of the code are printed.
The flags are as set in {\tt WUFLAG}, but the first letter is
changed to an {\tt N} standing for {\tt NCqed}.
Next, the relevant physical input constants are printed the meaning of
which is described in section~\ref{input}.
The next paragraph of output begins with the fermion pair indices.
The correspondence of fermions and indices is given in
table~\ref{fcodes}.
\begin{table}[bt]
\begin{center}
\vspace{1cm}
\caption{{\it Correspondence between fermion pair indices and fermion
  types for the} {\tt NCqed} {\it branch of} {\tt GENTLE/4fan}.
}
\label{fcodes}
\vspace{.25cm}
\begin{tabular}{|c||c|c|c|c|}
  \hline
  \raisebox{0.pt}[2.5ex][0.0ex]
  Fermion Pair Index & 1 & 2 & 3 & 4 \\ \hline
  \raisebox{0.pt}[2.5ex][0.0ex]
  Fermion Type       & $\nu_\mu, \nu_\tau$ & $\mu, \tau$ & $u, c$ &
                       $d, s, b$ \\ \hline
\end{tabular}
\end{center}
\vspace{.5cm}
\end{table}
Then, for the two fermion types used in the computation, the color
factors ({\tt RNCOU1}, {\tt RNCOU2}), the fermion masses ({\tt AM1},
{\tt AM2}), and the lower invariant mass cuts ({\tt CUTM12},
{\tt CUTM34}) are printed.
Closing this paragraph of output, the minimum value {\tt CUTXPR} of the
fraction $s'/s$ is printed.
Finally, together with the relative and absolute required precisions
{\tt EPS1} and {\tt ABS1}, the total \xsecs\ are output for each
required center of mass energy.
%
%
\subsection{Use of the program}
\label{use}
In this section, we will give a short guide to the compilation and use
of {\tt GENTLE/4fan}.

We recommend to exhaust all means of compiler optimization, because,
according to our experience, the CPU time consumption can thus be
reduced by a factor of two or more.
When available, the use of a compiler option that makes the default size
of floating-point constants REAL*8 may be advantageous.
For {\sf HP} computers, we recommend
\begin{verbatim}
  f77 +O4 +Onolimit +ppu -K -R8 gentle.f  .
\end{verbatim} 
For {\sf Silicon Graphics} computers, compilation with
\begin{verbatim}
  f77 -r8 -O3 -non_shared -G 32 -jmpopt -mips2 -static gentle.f
\end{verbatim}
has proved successful.

Unless the user wants to run {\tt GENTLE/4fan} with default flags, the user 
should set the flags described in section~\ref{input}.
The setting of a flag {\tt IFLAG} to a value {\tt IVALUE} is
accomplished by adding the line
\begin{verbatim}
  CALL WUFLAG('IFLAG',IVALUE)
\end{verbatim}
after the {\tt DATA} statements in the {\tt MAIN} of {\tt
  GENTLE/4fan}.
Please note that the numerical format of {\tt IVALUE} is {\tt INTEGER}.
Similarly, for {\tt IANO}=8, the anomalous couplings {\tt XG}, {\tt YG},
{\tt XZ}, {\tt YZ}, {\tt ZZ}, {\tt DG}, and {\tt DZ} may be set by calls like
\begin{verbatim}
  CALL WUFLAG('XG',VALUE)
\end{verbatim}
at the same place in the {\tt MAIN}.
Please note that the numerical format  of {\tt VALUE} is {\tt DOUBLE PRECISION}.
In addition, the physical input variables {\tt GFER}, {\tt ALPW},
{\tt AME}, {\tt AMW}, {\tt AMZ}, {\tt GAMZ}, and {\tt ALPHS} for the
{\tt CC} branch of the code as well as {\tt AMHIG} and {\tt GAMHIG} for
the {\tt 4fan} branch can be changed by analogous calls to {\tt WUFLAG}.
Finally, it is possible to change the physics input to the {\tt 4fan}
and {\tt NCqed} branches of the code in the subroutines {\tt bbmmin}
and {\tt NCIN00} respectively.

To determine the physics quantities and the degrees of moments to be
computed and printed, the user must set the range of the loops over
{\tt IREGIM} and {\tt IMOMN} (see page~\pageref{iremom}).

Cuts on the final state invariant pair masses may be applied in
the {\tt CC} and the {\tt 4fan} branches of the code by altering
{\tt S1MIN}, {\tt S1MAX}, {\tt S2MIN}, and {\tt S2MAX} in the
{\tt MAIN}.
In the {\tt NCqed} branch, final state invariant pair mass cuts are
implemented by setting {\tt X12MIN}, {\tt X12MAX}, {\tt X34MIN}, and
{\tt X34MAX} in the subroutines {\tt NCGENT} and
{\tt FINC34} (
{{\tt X12} $=\SONE/s$, {\tt X34} $=\STWO/s$}).
%
%
%
\section*{Acknowledgments}
\label{acknow}
We gratefully acknowledge the valuable contributions of M.~Bilenky in
early phases of the project. We benefitted greatly from discussions with
and extensive tests of the program by Peter Clarke, Douglas
Ferguson, Som Ganguli, Bingnian Jin, Martin Gr\"unewald. We would also
like to thank A.~Ballestrero, F.A.~Berends, M.~Dubinin, S.~Jadach,
M.~Moretti, O.~Nicrosini, T.~Ohl, G.~Passarino, R.~Pittau, M.~Pohl,
M.~Skrzypek, and Z.~Was for kind and productive collaboration during
the LEP~2 and LC workshops in 1995 and 1996. 
%
\clearpage
\appendix

%
%
\section{Numerical Integration in {\tt GENTLE/4fan}}
\label{integration}
\ezero
The integrations over $s_1,s_2$ are performed numerically by a
successive use of an adaptive one-dimensional Simpson integration routine.
One (two) more integrations are needed to calculate initial state
QED corrections in the flux function (structure function) approach.
Mappings of the integrands may be chosen with flag {\tt IMAP}. 
We now describe the mappings.

After the analytical integration over the 
six angular variables, one has to treat Breit-Wigner resonances
arising from the exchange of intermediate bosons:
\bq
\label{bw0map}
f_{BW}(V,s) \sim \frac{1}{(s-M_V^2)^2+\Gamma_V^2 M_V^2},\ \ \ V=Z,W,H.
\eq
In the case of neutral current processes, additional peaks appear 
at small momentum transfer due to virtual photons or gluons,
\bq
\label{ir0map}
f_{low}(V,s) \sim \frac{1}{s},\ \ \ V=\gamma,g.
\eq

For charged current processes, the standard mapping of the
Breit-Wigner resonance of the $W$ boson is applied in {\tt
GENTLE/4fan}.  

In the case of neutral current processes, the 
peaks at small momentum transfer and the
Breit-Wigner resonances of the $Z$ and the Higgs boson have to be
mapped simultaneously.  
In {\tt GENTLE/4fan}, the singularity at small momentum transfer
is mapped first by the usual
transformation to a new variable $\bar s$,
\bq
\label{irmap}
\bar s = f_1(s) = \ln s.
\eq
Then follows a mapping of the $Z$ resonance, which does not destroy 
the previous mapping,
\bq
\label{bwmap}
\bar{\bar s} = f_2(\bar s) = \bar s + \frac{c}{[\bar\Gamma_V\bar M_V]}
               \arctan\frac{\bar s-\bar M_V^2}{[\bar\Gamma_V\bar M_V]}.
\eq
The variables $\bar M_V^2$ and $[\bar\Gamma_V\bar M_V]$ are calculated
taking into account the previous mapping,
\bq
\bar M_V^2 = f_1(M_V^2),\ \ \ 
[\bar\Gamma_V\bar M_V] = f_1(M_V^2+\Gamma_VM_V)-f_1(M_V^2).
\eq
The function $f_2$ is inverted numerically by Newton's Method,
\bq
\bar s_{i+1} = \bar s_i - \frac{f_2(\bar s_i)}{f_2'(\bar s_i)},
\eq
which always converges for the starting value $\bar s_0=\bar M_V^2$.
The free constant $c$ in equation (\ref{bwmap}) is
optimized for every final state separately. 
However, the quality of the mapping
(\ref{bwmap}) is sensitive to the order of magnitude of $c$ only. 
In the case of Higgs production, a procedure similar to 
(\ref{bwmap}) is added to map the Higgs resonance.

\section{Anomalous Triple-Boson Couplings}
\label{anocoup}
\ezero
In addition to the \sm\ Lagrangian, we consider the following
{\sf CP} conserving operators of dimension 6
~\footnote[10]{
For definitions of anomalous couplings, see
e.g.~\cite{hagiwara0,LEP2ano}. 
}:
\begin{eqnarray}
  \Delta{\cal L}_{\mbox{eff}} & = &
  g'\frac{\alpha_{B\phi}}{m^2_W}\left(D_\mu\Phi\right)^\dagger
  B^{\mu\nu}\left(D_\nu\Phi\right) +
  g\frac{\alpha_{W\phi}}{m^2_W}\left(D_\nu\Phi\right)^\dagger
  {\overrightarrow{\tau}} {\overrightarrow{W}}^{\mu\nu} \left( D_{\nu}\Phi 
\right)
  \nl
 & & \mbox{} +
 g\frac{\alpha_{W}}{6m^2_W}\overrightarrow{W}^\mu_\nu
\left(\overrightarrow{W}^\nu_\rho
 \times \overrightarrow{W}^\rho_\mu\right) .
 \label{deltala}
\end{eqnarray}
In the unitary gauge, these operators lead to the following effective
Lagrangian for the $WWV$ vertex:
\begin{eqnarray}
  {\cal  L}^{WWV}_{\mbox{eff}} & = & {\rm i} g_{WWV} \left[\:
    \vphantom{\frac{\lambda_V}{m^2_W}}
    g^V_1 \left(W^+_{\mu\nu}W^{-\mu}-W^{+\mu}
    W^-_{\mu\nu}\right)V^\nu \right.\nl
  & & \hspace{1.8cm} \left.\mbox{}+\kappa_VW^+_\mu W^-_\nu V^{\mu\nu} +
      \frac{\lambda_V}{m^2_W} W^{+\nu}_\mu W^{-\rho}_\nu V^\mu_\rho
      \right]~,
  \label{anolag}
\end{eqnarray}
where $V\in\{Z,\gamma\}$.
Therefore, one has six relevant anomalous triple-boson couplings, namely
$g_1^\gamma$, $g_1^Z$, $\kappa_\gamma$, $\kappa_Z$, $\lambda_\gamma$,
and $\lambda_Z$.
Electromagnetic gauge invariance requires $g_1^\gamma=1$.
In the HISZ-scenario~\cite{hagiwara} ({\tt IANO}$=\pm7$), one sets
$\alpha_{B\phi}=\alpha_{W\phi}$.
Thus, in this scenario, one obtains relations between the parameters
$\kappa_\gamma$, $\kappa_Z$, and $g_1^Z$.

For the $WWZ$-vertex,
we also include a {\sf C} {\em and} {\sf P} violating term in
the Lagrangian: 
\begin{eqnarray*}
  {\cal L}_Z &=&\mbox{}-\frac{e z_Z}{m^2_W} \,
  \partial_\alpha\hat{Z}_{\rho\sigma}
  \left( W^{+\alpha}\stackrel{\leftrightarrow}{\partial^\rho}W^{-\sigma}-
  W^{+\sigma}\stackrel{\leftrightarrow}{\partial^\rho}W^{-\alpha}\right)~,
\end{eqnarray*}
where we have introduced the dual field strength tensor
\begin{eqnarray*}
  \hat{Z}_{\rho\sigma}&=&\frac{1}{2}\epsilon_{\rho\sigma\alpha\beta}
   Z^{\alpha\beta}~.
\end{eqnarray*}

To disentangle the anomalous couplings' contributions to the \xsec, we
use redefined parameters:
\begin{eqnarray}
\nl
\delta_Z&=&(g_1^Z-1)\cot\theta_W,
\label{arrbeg}
\\ \nl 
x_\gamma&=&\kappa_\gamma-1,
\\ \nl
x_Z&=&(\kappa_Z-1)(\cot\theta_W+\delta_Z),
\\ \nl 
y_\gamma&=&\lambda_\gamma,
 \\ \nl
y_Z&=&\lambda_Z\cot\theta_W,
\\ \nl 
z_Z&=&z_Z.
\label{arrend}
\end{eqnarray}
The {\tt GENTLE/4fan} user can set values for the anomalous couplings
defined in equations~(\ref{arrbeg}) to~(\ref{arrend}).
In the \sm, the anomalous couplings~(\ref{arrbeg}) to~(\ref{arrend})
vanish.
The special case of anomalous couplings induced by a heavy extra
neutral gauge boson is described in appendix~\ref{extra}. 
%
%
\section{Treatment of Extra Neutral Gauge Bosons}
\label{extra}
\ezero
%
The effect of an extra neutral gauge boson $Z'$ can be described with
{\tt GENTLE/\-4fan} by
two anomalous couplings $g^*_{WW\gamma}$ and $g^*_{WWZ}$ of the
photon and the Standard Model $Z$ boson to $W$ pairs \cite{pankov},
\bq
g^*_{WW\gamma}=1+\delta_\gamma \mbox{\ \ and\ \ }
g^*_{WWZ}=\cot\theta_W+\delta_Z,
\eq
where
\ba
\label{ze}
\delta_\gamma &=& 
g_{WWZ_1}\left(\frac{a_1}{a}-\frac{v_1}{v}\right)v(1+\Delta\chi)\chi_Z 
+ g_{WWZ_2}\left(\frac{a_2}{a} -\frac{v_2}{v}\right)v\chi_2,\nll
\delta_Z &=& -\cot\theta_W
+g_{WWZ_1}\frac{a_1}{a}(1+\Delta\chi) 
+ g_{WWZ_2}\frac{a_2}{a}\frac{\chi_2}{\chi_Z}, 
\ea
and
\bq
\label{chi}
\chi_Z=\frac{s}{s-M_Z^2},\ \ \ 
\Delta\chi=-\frac{2M_Z(M_Z-M_1)}{s-M_Z^2},\ \ \ 
\chi_2=\frac{s}{s-M_2^2}.\ \ \ 
\eq
In equations (\ref{ze}) and (\ref{chi}), $M_i,v_i,a_i,\ i=1,2$
denote the masses and electron couplings of the mass eigenstates $Z_1$
and $Z_2$ , which are, in general, a result of a mixing of the 
gauge symmetry eigenstates $Z$ and $Z'$,
\ba
\label{zepmix}
Z_1&=&Z\cos\theta_M + Z'\sin\theta_M,\ \ \ 
Z_2=-Z\sin\theta_M + Z'\cos\theta_M,\nll
v_1&=&v\cos\theta_M + v'\sin\theta_M,\ \ \ \ \,\,
v_2=-v\sin\theta_M + v'\cos\theta_M,\nll
a_1&=&a\cos\theta_M + a'\sin\theta_M,\ \ \ \ \,\,
a_2=-a\sin\theta_M + a'\cos\theta_M,
\ea
where $v,a\ (v',a')$ denote the couplings
of the symmetry eigenstates $Z\ (Z')$ to electrons and 
$\theta_M$ is the $ZZ'$ mixing angle.
The couplings $g_{WWZ_1}$ and $g_{WWZ_2}$ are fixed by the condition
that only the $Z$ couples to $W$ pairs,
\bq
g_{WWZ_1}=\cot\theta_W\cos\theta_M,\ \ \ 
g_{WWZ_2}=\cot\theta_W\sin\theta_M.
\eq
The mass shift from $M_Z$ to $M_1$ in the $Z_1$ propagator is absorbed
into $\delta_\gamma$ and $\delta_Z$ through $\Delta\chi$ assuming
$M_Z-M_1\ll M_Z$ and $M_Z-M_1\ll (s-M_Z^2)/(2M_Z)$. 
Both approximations are equivalent for $W$ pair production.

\section{Options of {\tt GENTLE}}
\label{feattab}
Table~\ref{gentab} shows the observables {\tt GENTLE} can calculate
and the options, which may be chosen.
Please note that in the {\tt CC11} family anomalous couplings are
included only for the {\tt CC03} process.

\begin{table}[bhtp]
\vspace{0.5cm}
\begin{center}
\caption{\it{Options of {\tt GENTLE};
SF - structure function approach;
FF - flux function approach;
h.o. - higher order QED corrections;
Coul. - Coulomb singularity.}
\label{gentab} 
}
\vspace{0.5cm}
\begin{tabular}{|l|c|c|c|c|c|c|c|c|c|}
\hline
&{\tt IREGIM}&on-shell&Born&SF&FF&h.o.&Coul.&non-uni.&anom.c.
\\\hline
\raisebox{0.pt}[2.5ex][0.0ex]{$\sigma^{\tt CC03}$}&0&+&+&+&+&+&+&+&+
\\\hline
\raisebox{0.pt}[2.5ex][0.0ex]{$\sigma^{\tt CC03}$}&1&+&--&+&+&+&+&+&+
\\\hline
\raisebox{0.pt}[2.5ex][0.0ex]{$\sigma^{\tt CC03}$}&2&+&--&+&--&+&+&--&+
\\\hline
\raisebox{0.pt}[2.5ex][0.0ex]{$\sigma^{\tt CC03}$}&3&+&--&+&+&+&+&+&+
\\\hline
\raisebox{0.pt}[2.5ex][0.0ex]{$\sigma^{\tt CC03}$}&4&+&+&+&--&+&+&--&--
\\\hline
 \raisebox{0.pt}[3.5ex][2.0ex]{$\frac{\displaystyle {\rm d} \sigma^{\tt
      CC03}}{\displaystyle {\rm 
d}\cos\!\theta}$}&--&+&+&+&--&+&+&--&+
\\\hline
\raisebox{0.pt}[2.5ex][0.0ex]{$\sigma^{\tt CC11}$}&0&--&+&+&+&+&+&--&+
\\\hline
\raisebox{0.pt}[2.5ex][0.0ex]{$\sigma^{\tt CC11}$}&1&--&--&+&+&+&+&--&+
\\\hline
\raisebox{0.pt}[2.5ex][0.0ex]{$\sigma^{\tt CC11}$}&2&--&--&+&--&+&+&--&+
\\\hline
\raisebox{0.pt}[2.5ex][0.0ex]{$\sigma^{\tt CC11}$}&3&--&--&+&+&+&+&--&+
\\\hline
\raisebox{0.pt}[2.5ex][0.0ex]{$\sigma^{\tt NC02}$}&--&--&+&--&+&+&--&+&--
\\\hline
\raisebox{0.pt}[2.5ex][0.0ex]{$\sigma^{\tt NC08}$}&--&--&+&--&+&+&--&+&--
\\\hline
\raisebox{0.pt}[2.5ex][0.0ex]{$\sigma^{\tt NC32}$}&--&--&+&+&+&+&--&--&--
\\\hline
\end{tabular}
\end{center}
\end{table}

\newpage
%

\end{document}